\documentclass[11pt,english]{article}
\usepackage[latin9]{inputenc}
\usepackage{lmodern}
\usepackage{geometry}
\usepackage{color}
\usepackage{hyperref}
\usepackage{url}
\usepackage{titling}
\usepackage{amsbsy}
\usepackage{babel}
\usepackage{amssymb}
\usepackage{amsmath}
\usepackage{multirow}
\usepackage{mathtools}
\usepackage{esint}
\usepackage[authoryear]{natbib}
\usepackage{titletoc}
\usepackage{booktabs}
\usepackage{multirow}
\usepackage{subcaption}
\usepackage{amsmath}
\usepackage{hyperref}
\usepackage{tikz}
\usepackage{setspace}
\usepackage{comment}
\usepackage{color}
\usepackage{colortbl}
\usepackage{amssymb}
\usepackage{pifont}
\usepackage{color, colortbl}
\definecolor{Gray}{gray}{0.9}

\newcommand*{\DataPath}{Figures}

\newcommand*{\FigurePath}{Figures}

\newtheorem{model}{\modelname}

\providecommand{\modelname}{Model}

\title{Measuring Consumer Sensitivity to Audio Advertising:\\ A Long-Run  Field Experiment on Pandora Internet Radio\thanks{\footnotesize We wish to recognize Chris Irwin and Jack Krawczyk for their leadership and vision in getting this experiment implemented. We are grateful to Pandora for allowing us to present these results to the scientific community; our agreement with management is to present our scientific results without discussing policy implications for the company. We thank Neil Buckman, Alexa Maturana-Lowe, Isaac Hill, Ashley Bischof, Carson Forter, and Puya Vahabi for considerable help in producing and organizing the data. We also thank Andrew Asman, Oliver Bembom, Dean Eckles, Gautam Gowrisankaran, David Hardtke, Michael Hochster, Garrett Johnson, Sam Lendle, Randall Lewis, Jura Liaukonyte, Adam McCloskey, Steve Tadelis, Mo Xiao, Hongkai Zhang, Zhen Zhu, and seminar participants at Brown University, UC Berkeley, Columbia University Business School, the Central European Institute, and the University of Tennessee for helpful comments. }}

\author{ Ali Goli \\ \textit{University of Washington} \and Jason Huang \\ \textit{Databricks} \and David Reiley \\ \textit{Sirius XM Pandora} \and Nickolai M. Riabov  \\ \textit{Airbnb}}

\date{}

\usepackage{natbib} 

\begin{document}
\maketitle
\vspace{-40pt}
\begin{center}
\textit{First version: April 21, 2018}

\textit{This version: December 4, 2024}
\end{center}
\vspace{5pt}
\begin{abstract}
A randomized experiment with almost 35 million Pandora listeners enables us to measure the sensitivity of consumers to advertising, an important topic of study in the era of ad-supported digital content provision. The experiment randomized listeners into nine treatment groups, each of which received a different level of audio advertising interrupting their music listening, with the highest treatment group receiving more than twice as many ads as the lowest treatment group. By maintaining consistent treatment assignment for 21 months, we measure long-run demand effects and find ad-load sensitivity three times greater than what we would have obtained from a month-long experiment. We show the negative impact on the number of hours listened, days listened, and probability of listening at all in the final month. Using an experimental design that separately varies the number of commercial interruptions per hour and the number of ads per commercial interruption, we find that listeners primarily respond to the total number of ads per hour, with a slight preference for more frequent but shorter ad breaks. Lastly, we find that increased ad load led to an increase in the number of paid ad-free subscriptions to Pandora. Importantly, we show that observational methods often lead to biased or even directionally incorrect estimates of these effects, highlighting the value of experimental data.

\end{abstract}

\pagebreak

\doublespacing
\section{Introduction} \label{sect:intro}

Two-sided markets have emerged as a cornerstone of the digital economy, fundamentally reshaping how businesses connect consumers and service providers. From e-commerce platforms like eBay to ride-sharing services like Uber, these markets have transformed numerous industries. Among these, advertising-supported digital content platforms represent a particularly significant and complex class of two-sided markets~\citep{Rysman2009}. Platforms such as Yahoo, Google, Facebook, YouTube, CNN.com, Pandora, Spotify, and Hulu exemplify how free, ad-supported content has become integral to the modern digital landscape. Like other two-sided markets, these platforms must carefully balance the competing interests of advertisers and consumers.

The quantity of advertising acts as an implicit price that consumers pay for access to free content, a concept explored extensively by \citet{anderson2006media}. While consumers generally find advertising disruptive \citep{wilbur2008two,goldstein2014economic}, it is the cost they bear for content consumption. This creates a delicate balance for platforms, as they must determine the optimal level of advertising that maximizes revenue without driving away users \citep{armstrong2006competition,rochet2006two}. Given that ad load can be an endogenous outcome, dependent on user behavior and actions as well as supply factors, it is crucial to experimentally shift this variable to understand its causal effect on consumption. However, large-scale ad-load experiments have been uncommon. This scarcity may be partly attributed to the historical dominance of traditional ad-supported broadcast media like television and radio, where technological constraints made it much harder to experiment than in the digital world. While some traditional media could conduct geo-testing, this approach often suffered from limited statistical power due to the smaller number of geographic areas available for experimentation. \citet{lodish1995tv} presented an innovative experiment that substituted one ad for another in a broadcast feed, but even that technology was limited to thousands (rather than millions) of experimental observations, which limited statistical power, and the number of ads in their experiments remained fixed due to constraints of the scheduled TV programming. 

The rise of digital platforms has created unprecedented opportunities for comprehensive experimentation. While field experiments in economics have become increasingly common since early examples at the turn of the century \citep{lucking1999using, list2000demand, list2002effects}, their scope and scale have expanded dramatically. Early experiments, particularly in development economics \citep{duflo2006field, banerjee2020field}, often employed binary pricing tests that revealed significant effects when changing from zero to a positive price. For instance, \citet{MiguelEtAl2002} studied free versus \$0.30 deworming medicine, while \citet{CohenDupas2007} examined free versus \$0.60 mosquito nets. The digital revolution has extended experimentation far beyond development economics, making it central to business strategy across various domains. Companies now routinely run experiments to measure advertising effectiveness \citep{johnson2017less,LewisReiley2014}, optimize paid search advertising \citep{BlakeNoskoTadelis2015}, improve search result rankings \citep{ursu2018power}, test personalized email content \citep{sahni2018personalization}, refine user interface design \citep{kohavi2017surprising}, evaluate promotions \citep{hitsch2024heterogeneous}, and develop pricing strategies \citep{DupeMisra_2017}.

The value of experimentation becomes particularly apparent when comparing experimental results to observational studies. \citet{gordon2019comparison,gordon2023close} demonstrate that experimental methods often yield substantially different results compared to observational methods in online advertising effectiveness measurement. This divergence highlights the potential for misleading conclusions when relying solely on observational data, especially in complex digital ecosystems where user behavior and platform dynamics are intertwined.

However, many business experiments fall short of ideal scientific standards due to three key limitations: short durations, small sample sizes, and narrow variation in decision variables. These constraints severely limit firms' ability to detect small treatment effects, explore effect heterogeneity reliably, or generalize results across broader decision spaces. Due to resource constraints and pressure for quick results, many experiments run for days or weeks rather than months or years, often with limited participants and restricted variable ranges. Many firms consider an experiment ``long-run'' if it merely achieves statistical significance, rather than allowing sufficient time for consumer behavior to reach equilibrium.

This approach is particularly problematic for treatments whose effects may take time to manifest fully. For instance, consumers may need time to perceive changes in ad load, especially when these changes are subtle compared to easily observable price changes. Moreover, for habit-forming goods like music streaming services, changes in habit stock can alter the marginal utility of current consumption, leading to gradual behavioral adjustments \citep{becker1988theory}. This dynamic process involves both learning and new habit formation, potentially impacting long-term treatment effects.

Furthermore, short-run experimentation with small samples can lead to high false discovery rates, potentially misleading decision-makers about intervention effects \citep{camerer2018evaluating,berman2022false}. This is especially problematic when examining changes to core product features or user experience elements that act as implicit costs, such as search result rankings, product assortments, recommendation algorithms, or user interface designs. Consumer responses to such changes often evolve over extended periods as they reassess usage patterns and explore alternatives.

Our study aims to fill this gap by conducting what we believe to be the first large-scale, long-term randomized experiment on advertising load in a digital platform. We varied the number of ads for 35 million Pandora listeners over a period of 21 months from 2014 to 2016. This extended timeframe is crucial, as we find that the long-run elasticity of demand with respect to ad load is three times larger than the short-run elasticity. We experimented with nine different levels of advertising intensity, with the highest group receiving twice as many ads as the lowest group. This wide range allowed us to trace out a continuous demand curve for Pandora's music service, treating the quantity of advertising as the ``price'' paid by consumers. By imposing consistent treatment assignment throughout the 21-month period, we ensured that we measured long-run rather than merely short-run consumer demand. This approach allowed us to measure the long-run effects of advertising intensity on various outcomes, most notably the total quantity of music consumed by listeners, building on earlier work on ad avoidance by consumers \citep{wilbur2008two,anderson2011platform}. 

To summarize, this research design offers several important advantages:

\begin{itemize}
\item[\textbf{1.}] \textbf{Randomization:} The randomized nature of our experiment allows us to measure true causal effects, avoiding the potential confounds and untestable assumptions often present in observational studies.
\item[\textbf{2.}] \textbf{Range of Ad Loads:} By varying advertising loads across a wide range (with the highest group receiving twice as many ads as the lowest), we can observe the relationship between ad quantity and listening behavior across a substantial price range. This enables us to estimate a more complete relationship between ad quantity and listening behavior.
\item[\textbf{3.}] \textbf{Long-term Measurement:} The 21-month duration enables us to capture long-run effects, which prove to be significantly larger than short-run impacts. This addresses concerns about the external validity of short-run experimental results.
\item[\textbf{4.}] \textbf{Large Sample Size:} Our massive sample size of 35 million consumers provides precise estimates of treatment effects, allowing us to detect even subtle changes in consumer behavior and explore heterogeneous responses across different user segments and ad scheduling methods.
\end{itemize}

Our study yields important insights into optimal ad scheduling and the relationship between free and paid versions of the service. We find that, conditional on holding the ad load fixed, consumers prefer more frequent but shorter ad breaks, contributing to the literature on optimal ad load and scheduling~\citep{wilbur2008two,wilbur2013correcting,kar2015selection,rafieian2023optimizing}. Moreover, our results reveal a significant increase in paid subscriptions as ad load increases, highlighting the complex interactions in freemium models and adding to the literature on versioning in information goods~\citep{shapiro1998versioning,bhargava2008research}.

To demonstrate the value of our experimental approach, we compare our results to those obtained using various observational methods, including panel data techniques, instrumental variables estimation, and the inclusion of granular fixed effects. We find that these methods often lead to biased estimates of the causal effects of ad load on listener behavior, sometimes even yielding incorrect directional effects. Crucially, we demonstrate that relying on observational data would have led to misleading conclusions both on ad scheduling and the relationship between the ad-supported and paid versions of the service. Observational methods led to inconclusive results on ad scheduling and suggested that the paid and ad-supported versions were complements, that is, increasing ad load would decrease ad-supported usage as well as subscriptions. These discrepancies underscore the importance of experimental data in accurately understanding platform dynamics and making informed business decisions.

There are a number of related studies on ad load we wish to highlight here. Several studies have examined the effects of advertising load across various contexts. \citet{BerryWaldfogel1999} consider advertising volume as a quality attribute for radio stations but do not measure this attribute directly, instead modeling it as a latent, unobserved quantity. In the television industry, \citet{wilbur2008two} uses observational data to study viewer responses to advertising, reporting a notably high audience advertising elasticity of about 2.5. His study suggests that viewers are generally averse to advertising, with a 10\% decrease in ad time potentially leading to a 25\% audience gain.  Interestingly, our findings indicate that observational methods in our context significantly overstate the effect of advertising on demand compared to causal effects from experimental variation, highlighting the value that experimental approaches offer in the digital world.

In the digital realm, \citet{sahniQME2015} finds no change in consumer usage of a restaurant-search website when slightly reducing ad load. By contrast, \citet{sahni2024consumers} document increased search engine usage with more prominent advertising, suggesting ads can complement search results by providing valuable information. \citet{HohnholdEtAl2015} examine ad blindness in search results, finding that increased ad quantity makes users less likely to click on ads over time. Relatedly, \citet{goldstein2014economic} conduct a field experiment to estimate the economic and cognitive costs of annoying display advertisements. They find that in plausible scenarios, running annoying ads can cost publishers more money than it earns. Further, \citet{moshary2024sponsored} provides empirical evidence that while sponsored search ads can steal business from organic listings and reduce the total volume of transactions, they still prove profitable for the platform due to the revenue from ads outweighing the reduction in commissions from transactions.

Our study extends this line of research with several unique strengths. We conduct an unprecedented large-scale, long-term experiment on ad load effects in music streaming, addressing limitations of previous studies. The experiment's scale (35 million listeners), duration (21 months), and wide range of ad loads (9 different treatments spanning a doubling of the number of ads) allow us to capture long-run effects and trace a comprehensive demand curve. Notably, we find that long-run elasticity of demand is significantly larger than short-run estimates, emphasizing the importance of extended observation periods in understanding ad-load impacts. Building on the work in the present paper, \citet{goli2021personalized} use a later experiment on Pandora to explore personalizing ad load to optimize subscription and ad revenues. They find that reallocating ads across users can enhance subscription profits without decreasing overall advertising profits, though this may come at a cost to consumer welfare. While they replicate our finding that it takes considerable time to reach steady state in response to ad load changes, their focus is on developing and evaluating personalized ad load policies using machine learning methods. Our paper's unique contributions include documenting the differences between long-run and short-run elasticities, analyzing how different margins of consumption respond, studying pod length versus frequency effects, and demonstrating the limitations of observational methods -- aspects not explored in their work. Together, these papers provide complementary insights into both the general effects of ad load and the potential for targeted strategies in managing ad-supported content platforms.

The remainder of this paper is structured as follows. Section \ref{sect:design} describes the institutional details of Pandora advertising and outlines our experimental design, including validation of the randomized treatment assignment. Section \ref{sec:measurement} presents our main results on listener sensitivity to advertising load, focusing on outcomes such as hours listened, days listened, and probability of listening at all, using experimental variation. In Section \ref{sec:observational}, we demonstrate that using observational methods such as OLS, OLS with fixed effects, panel, or instrumental variable regressions would yield misleading results, highlighting the importance of experimental data. Section \ref{sec:business} explores some business implications of our findings, including the effects of ad scheduling on listener activity and the impact of ad load on substitution to the subscription service. We also examine heterogeneous effects of ad load on subscription and churn rates across age groups, which can be leveraged to personalize ad load and optimize user experience and revenues. The paper concludes in Section \ref{sect:conclusion} with a discussion of our results and their broader implications for digital platforms and two-sided markets. Throughout, we emphasize the value of long-term, large-scale experimentation in uncovering causal relationships and informing business decisions.

\section{Experimental Design and Background} \label{sect:design}

Between June 2014 and April 2016, Pandora Internet Radio conducted a large-scale experiment to measure the sensitivity of listeners to the number of audio ads they hear interrupting their music stream. During this time period, we randomized 19\% of listeners into one of ten different treatment groups: nine treatment groups each with 1\% of listeners, plus a 10\% control group receiving the Pandora status quo. Each treatment group received different levels of audio advertising when they listened to Pandora via their mobile devices, with an individual's treatment assignment kept consistent for the entire period of the experiment. New listeners joining Pandora were assigned to treatment using the same randomization (a hashing function of user ID with a unique randomization seed not used in any other current or historical experiments at Pandora). This experimental design allows us to measure the effect of ad load increases in equilibrium, which includes the impact on both existing and new users, as new users are added to all conditions with their respective probabilities.

We implemented a 3x3 experimental design, separately varying the number of commercial interruptions per hour and the number of ads per commercial interruption.  At the time the experiment began, the status quo was for listeners to receive four commercial interruptions per hour. We will often refer to each commercial interruption as a ``pod'' of ads.  The status quo at the start of the experiment was for pods to alternate deterministically between one and two ads per interruption, for an average of 1.5 ads per pod. The experiment assigned listeners to receive either 3, 4, or 6 interruptions per hour, and to receive either 1, 1.5, or 2 ads per interruption, with each of the 9 possible combinations getting its own 1\% group of listeners.  We will use the following shorthand to refer to these treatments: 3x1 for three 1-ad pods per hour, 6x1.5 for six pods per hour alternating between one and two ads per pod, and so on. We note that the large 10\% control group was redundant with the status-quo 4x1.5 treatment. By varying ad load in these two separate dimensions, we enable ourselves to measure whether listeners have a preference for more commercial interruptions versus longer commercial interruptions, conditional on the platform's desired number of ads per hour.

Pandora's ad delivery system uses timers to control when listeners receive advertisements. These timers establish regular intervals at which listeners become eligible for ad breaks, with the specific timing varying by experimental condition. The system manages both the frequency of interruptions (controlled by the timer settings) and the number of ads played during each break (the ad pod length). For example, listeners in the 4x2 condition become eligible for an ad break every 15 minutes, while those in the 3x1 condition wait 20 minutes between potential ad breaks. The most frequent interruption schedule occurs in the 6x2 condition, where listeners become eligible for ads every 10 minutes. Once an ad pod is served, the timer resets, and the system checks at the end of each track whether the user is eligible for another ad pod.\footnote{End of a track is either when the track ends, or the user interrupts it by skipping the song.} Since ads are only served at the end of a track, the exact number of ads delivered can vary depending on the length of songs and the timing of listening sessions.

Additionally, due to resource constraints, Pandora chose to implement the experiment only in its mobile application (Android and iOS), not on the Pandora website or other pieces of client software, such as those in automobiles. Listening on mobile devices constituted approximately 80\% of Pandora listening during this time period, so the experiment does generate considerable variation in advertising exposure to individuals. However, listeners who listened only on the website from their desktop computers would receive no effective differences in treatment.\footnote{We acknowledge that this approach may lead to some attenuation of our elasticity estimates. See Section~\ref{sect:conclusion} for a discussion of this limitation.} The total amount of realized experimental treatment, therefore, depends on how much listening an individual does on different devices. In our analysis, we consider the impact on total listening across all devices, rather than restricting attention to listening on mobile devices, since many consumers listen via multiple devices, and we would expect listener perceptions of Pandora's value to depend on the total amount of advertising they receive during all listening sessions they do.

Pandora's ad delivery also depends on advertiser demand. Unlike in many digital advertising markets, there is no well-developed auction market for online audio advertising. Instead, all audio ads on Pandora are sold via forward contracts with advertisers. These contracts generally specify targeting attributes (such as ``males aged 25-44 who live in San Diego''), as well as frequency caps that prevent the same listener from hearing the same advertisement more than a specified number of times per day or week. Given these delivery constraints, Pandora sometimes has no appropriate ad to play for a given listener on a given scheduled ad-delivery opportunity. At such moments, the ad opportunity will go unfilled, and the listener will get to listen to more music instead of the scheduled advertisement. The realized ``fill rate'' (fraction of Pandora ad-delivery opportunities that actually get filled with an ad) can vary with listener attributes, with the time of day, and with the time of year.

Thus, we observe a considerable difference between the intended ad load and the realized ad load for each listener, with the latter being what we expect to affect listener behavior.  In our analysis, we therefore use instrumental-variables estimation, as treatment assignment causes plenty of exogenous variation in realized ad load, amidst other potentially endogenous variation.

To analyze listener response to advertising, we examine several key outcome measures in this experiment. Our primary metrics include total hours listened (combining both ad-supported and subscription listening for each user), number of days active in a given month, and probability of listening at all to Pandora during a given time period. We consider a user to be active in a given month if they have non-zero consumption hours across either the ad-supported or paid version of Pandora.\footnote{Unless otherwise stated, all our regressions and analyses in subsequent sections calculate activity and hours at the monthly level.} We also measure the impact on the probability of purchasing an ad-free subscription to Pandora, which was the ad-free counterpart to the ad-supported service and cost \$4.99 per month. For confidentiality reasons, we normalize the listening hours and days metrics by dividing all observations by the control-group mean and multiplying by 100.\footnote{The normalization by control group mean has minimal impact on our standard errors. As shown in Online Appendix C, the adjusted standard errors change by less than 1\% of our estimates' magnitude due to our large control group of 18 million users.} We use the abbreviation ``norm.'' in our tables to denote this normalization.

To validate our randomized experiment, we examine whether treatment assignment is balanced across pre-treatment characteristics. We focus on our key outcome variables measured in May 2014, the month before the experiment began. Table \ref{table:firstmon-stats} presents the means of these variables across all treatment conditions. Each observation represents a listener who used the ad-supported version of Pandora at least once during the experiment period. The means appear similar across groups, and formal $\chi^2$ tests comparing all ten treatment groups (nine treatment groups plus control, where control receives the same treatment as the 4x1.5 group) fail to reject the null hypothesis of equality at the 5\% level for each variable, as shown in Table \ref{table:firstmon-chisq}. These results confirm that our randomization was successful in creating balanced treatment groups.

\begin{table}[ht]
\centering
\caption{Summary of pre-treatment variables across treatment groups (means and standard errors). For data confidentiality, outcomes are normalized relative to their value in the control condition, which is why the control group always shows 100 for normalized variables.}
\label{table:firstmon-stats}
\resizebox{\textwidth}{!}{%
\begin{tabular}{lcccccccccc}
  \\ 
 \hline
Condition & 3x1 & 3x1.5 & 3x2 & 4x1 & 4x1.5 & 4x2 & 6x1 & 6x1.5 & 6x2 & Control \\ 
  \hline
  \hline
\rowcolor{gray!30} Total Hours (norm.) & 100.042 & 100.243 & 100.211 & 100.153 & 100.197 & 99.954 & 99.835 & 99.881 & 100.034 & 100.000 \\ 
\rowcolor{gray!30}     & (0.219) & (0.218) & (0.219) & (0.219) & (0.217) & (0.223) & (0.228) & (0.216) & (0.222) & (0.071) \\ 
  Days Active (norm.) & 99.955 & 100.093 & 100.201 & 100.032 & 100.079 & 99.952 & 99.996 & 99.941 & 99.949 & 100.000 \\ 
     & (0.139) & (0.139) & (0.139) & (0.139) & (0.139) & (0.138) & (0.139) & (0.139) & (0.139) & (0.044) \\ 
\rowcolor{gray!30} Audio Ads (norm.) & 100.061 & 100.251 & 100.271 & 99.868 & 100.086 & 99.830 & 99.607 & 99.827 & 99.939 & 100.000 \\ 
\rowcolor{gray!30}      & (0.230) & (0.231) & (0.230) & (0.230) & (0.230) & (0.230) & (0.228) & (0.229) & (0.230) & (0.073) \\ 
  Audio Pods (norm.) & 100.027 & 100.272 & 100.250 & 99.856 & 100.071 & 99.849 & 99.660 & 99.848 & 99.958 & 100.000 \\ 
       & (0.225) & (0.226) & (0.225) & (0.225) & (0.225) & (0.226) & (0.224) & (0.225) & (0.226) & (0.071) \\ 
\rowcolor{gray!30} Ad Capacity (norm.) & 100.088 & 100.213 & 100.351 & 99.993 & 100.032 & 99.984 & 99.790 & 99.893 & 99.947 & 100.000 \\ 
\rowcolor{gray!30}        & (0.218) & (0.218) & (0.222) & (0.218) & (0.217) & (0.218) & (0.217) & (0.218) & (0.218) & (0.069) \\ 
  Percent Paid Users & 1.101 & 1.109 & 1.098 & 1.097 & 1.108 & 1.083 & 1.100 & 1.096 & 1.108 & 1.103 \\ 
         & (0.008) & (0.008) & (0.008) & (0.008) & (0.008) & (0.008) & (0.008) & (0.008) & (0.008) & (0.002) \\ 
\rowcolor{gray!30} Percent Male & 46.609 & 46.644 & 46.700 & 46.648 & 46.698 & 46.668 & 46.718 & 46.689 & 46.623 & 46.671 \\ 
\rowcolor{gray!30}          & (0.037) & (0.037) & (0.037) & (0.037) & (0.037) & (0.037) & (0.037) & (0.037) & (0.037) & (0.012) \\ 
  Sample Size & 1,809,261 & 1,810,630 & 1,810,172 & 1,810,526 & 1,813,548 & 1,812,174 & 1,809,930 & 1,810,193 & 1,807,270 & 18,097,258 \\ 
   \hline
\end{tabular}%
}
\end{table}

\begin{table}[ht]
\centering
\caption{$\chi^2$ test for equality of means across all treatment groups, pretreatment period.}
\label{table:firstmon-chisq}
\begin{tabular}{lcc}
  \\ 
 \hline
 & Test Stat. & P Value \\ 
  \hline
   \hline
Total Hours & 3.972 & 0.913 \\ 
  Days Active & 3.367 & 0.948 \\ 
  Audio Ads & 7.172 & 0.619 \\ 
  Ad Capacity & 4.865 & 0.846 \\ 
  Male & 8.277 & 0.507 \\ 
  Paid Users & 10.050 & 0.346 \\ 
   \hline
\end{tabular}
\end{table}

Having validated the randomization, we now examine the realized treatment intensity across groups. Table \ref{table:tp-stats} shows the means of the amount of treatment received between June 2014 and March 2016\footnote{Technically, the experiment ended on April 7, 2016. For expositional convenience, our outcome period will be the ``final month'' of the experiment, defined as March 8 to April 7, 2016.} across all treatment groups. We see that treatment assignment does, as intended, manipulate the realized ad load. The highest treatment group receives 36\% more ads per pod and 33\% more pods per hour than the lowest treatment group, for a total of 84\% more ads per hour. For reference, the third row of the table shows the mean ad capacity per hour, or the number of audio ads that listeners would have received if every ad opportunity were filled by an advertiser. These numbers differ from the intended ad load numbers (6x2=12, 4x1.5=6, and 3x1=3) for several reasons having to do with the details of ad serving on Pandora. For example, listening sessions of different lengths can result in different numbers of ads per hour due to the details of the timing of ads. Another important reason for the differences is that the experiment manipulated advertising only for those listening via mobile apps, not for those listening via the Pandora website.

\begin{table}[ht]
\centering
\caption{Summary of realized advertising load during treatment period across experiment groups.}
\label{table:tp-stats}
\resizebox{\textwidth}{!}{%
\begin{tabular}{lcccccccccc}
  \\ 
 \hline
 & 3x1 & 3x1.5 & 3x2 & 4x1 & 4x1.5 & 4x2 & 6x1 & 6x1.5 & 6x2 & Control \\ 
  \hline
  \hline
\rowcolor{gray!30} Audio Ads per Hour & 2.735 & 3.184 & 3.735 & 3.083 & 3.625 & 4.254 & 3.620 & 4.293 & 5.021 & 3.628 \\ 
\rowcolor{gray!30}     & (0.001) & (0.002) & (0.002) & (0.001) & (0.002) & (0.002) & (0.002) & (0.002) & (0.002) & (0.001) \\ 
  Audio Ad Pods per Hour & 2.648 & 2.654 & 2.660 & 2.991 & 2.964 & 3.007 & 3.519 & 3.522 & 3.525 & 3.002 \\ 
     & (0.001) & (0.001) & (0.001) & (0.001) & (0.001) & (0.001) & (0.002) & (0.002) & (0.002) & (0.000) \\ 
\rowcolor{gray!30} Ad Capacity per Hour & 3.602 & 4.615 & 5.847 & 4.092 & 5.292 & 6.798 & 4.803 & 6.341 & 8.131 & 5.273 \\ 
\rowcolor{gray!30}      & (0.009) & (0.018) & (0.021) & (0.023) & (0.014) & (0.073) & (0.007) & (0.013) & (0.032) & (0.005) \\ 
  Percent with Non-zero Pods & 93.242 & 93.321 & 93.397 & 93.338 & 93.374 & 93.490 & 93.491 & 93.512 & 93.575 & 93.405 \\ 
       & (0.019) & (0.019) & (0.018) & (0.019) & (0.018) & (0.018) & (0.018) & (0.018) & (0.018) & (0.006) \\ 
\rowcolor{gray!30} Audio Ads per Pod & 1.054 & 1.224 & 1.430 & 1.054 & 1.237 & 1.430 & 1.055 & 1.234 & 1.430 & 1.229 \\ 
\rowcolor{gray!30}        & (0.000) & (0.001) & (0.001) & (0.000) & (0.001) & (0.001) & (0.001) & (0.001) & (0.002) & (0.000) \\ 
  Sample Size & 1,809,261 & 1,810,630 & 1,810,172 & 1,810,526 & 1,813,548 & 1,812,174 & 1,809,930 & 1,810,193 & 1,807,270 & 18,097,258 \\ 
   \hline
\end{tabular}%
}
\end{table}

\begin{figure}[!htbp]
\begin{center}
\includegraphics[scale = 0.45]{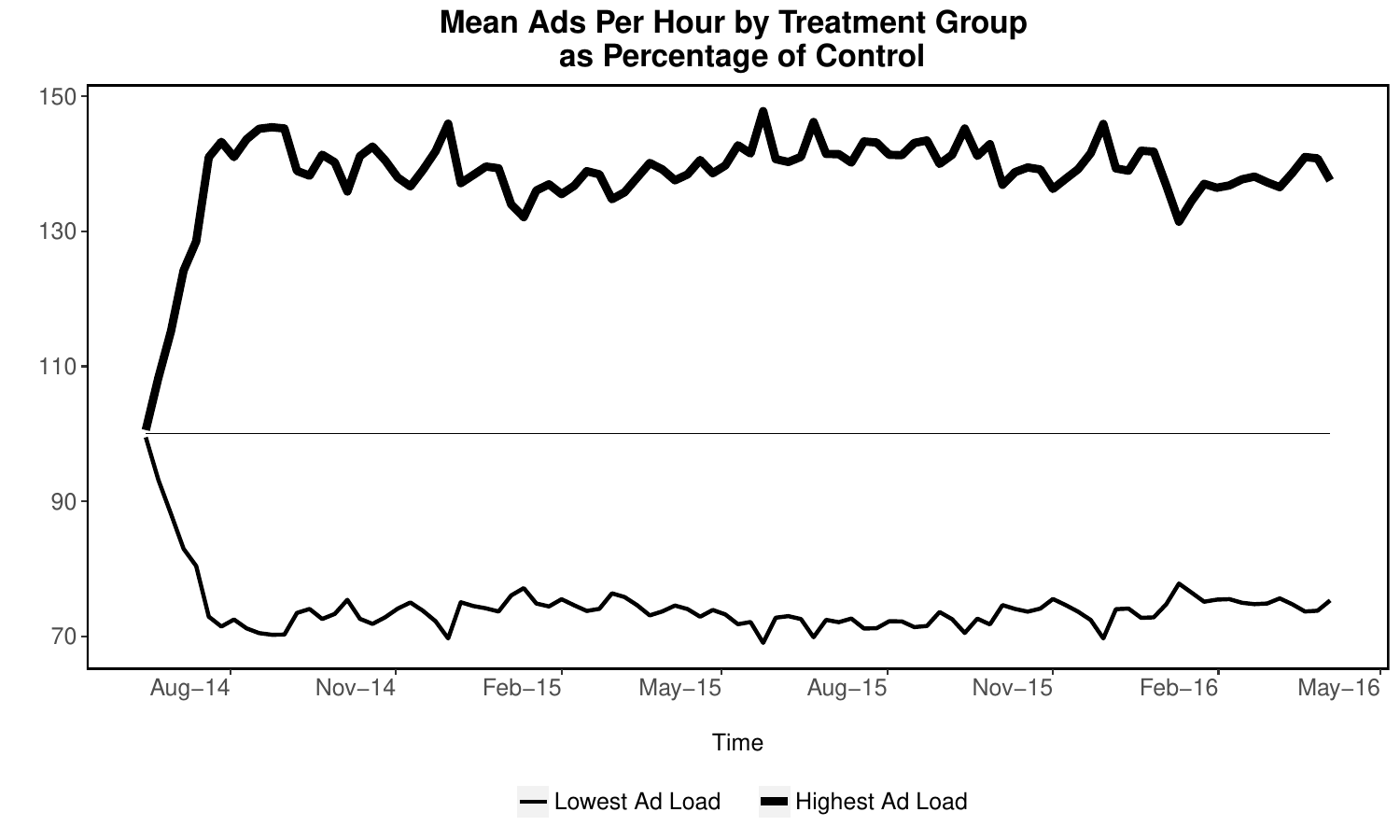}
\end{center}
\caption{Realized mean ad load per hour for highest (6x2) and lowest (3x1) treatment groups over the course of the experiment, normalized relative to the control group.}
\label{fig:ad-load-intertemp}
\end{figure}

Figure \ref{fig:ad-load-intertemp} shows how the amount of treatment varies from one week to the next. This time series plot shows the amount of ad-load treatment received by the highest (6x2) and lowest (3x1) treatment groups, divided by the ad load of the control group. We point out two key features. First, the experiment was designed to ramp up slowly over a period of six weeks, visible at the left side of the graph, in case listeners would find a sudden increase in ads to be jarring. Second, the amount of treatment-control difference varies somewhat over time, since realized ad load differs from intended ad load in a way that depends on advertiser demand. In particular, we can see that the treatment intensity is a bit higher in December than it is in January, because advertiser demand is very high in December and relatively low in January, which makes it easier to fill the ad capacity during this peak demand period.

\begin{figure}[htbp]
\begin{center}
\includegraphics[scale = 0.5]{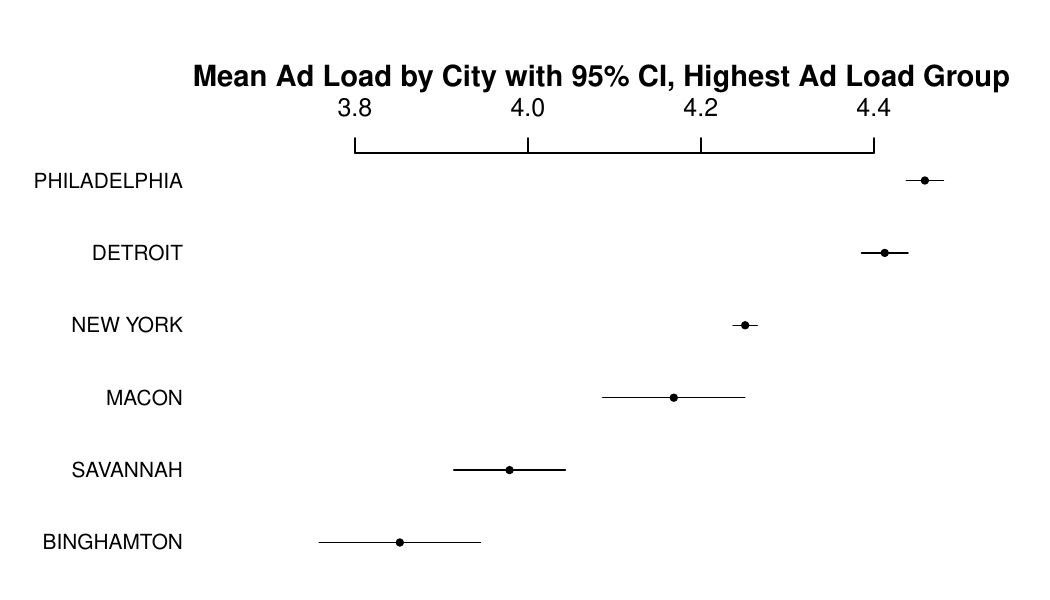}
\end{center}
\caption{Mean final month ad load by place of residence for six randomly selected cities within the highest intended treatment group. This figure illustrates the geographic variation in realized ad load even within a single treatment group.}
\label{fig:TG7-dmadiff}
\end{figure}

\begin{figure}[htbp]
\begin{center}
\includegraphics[scale = 0.5]{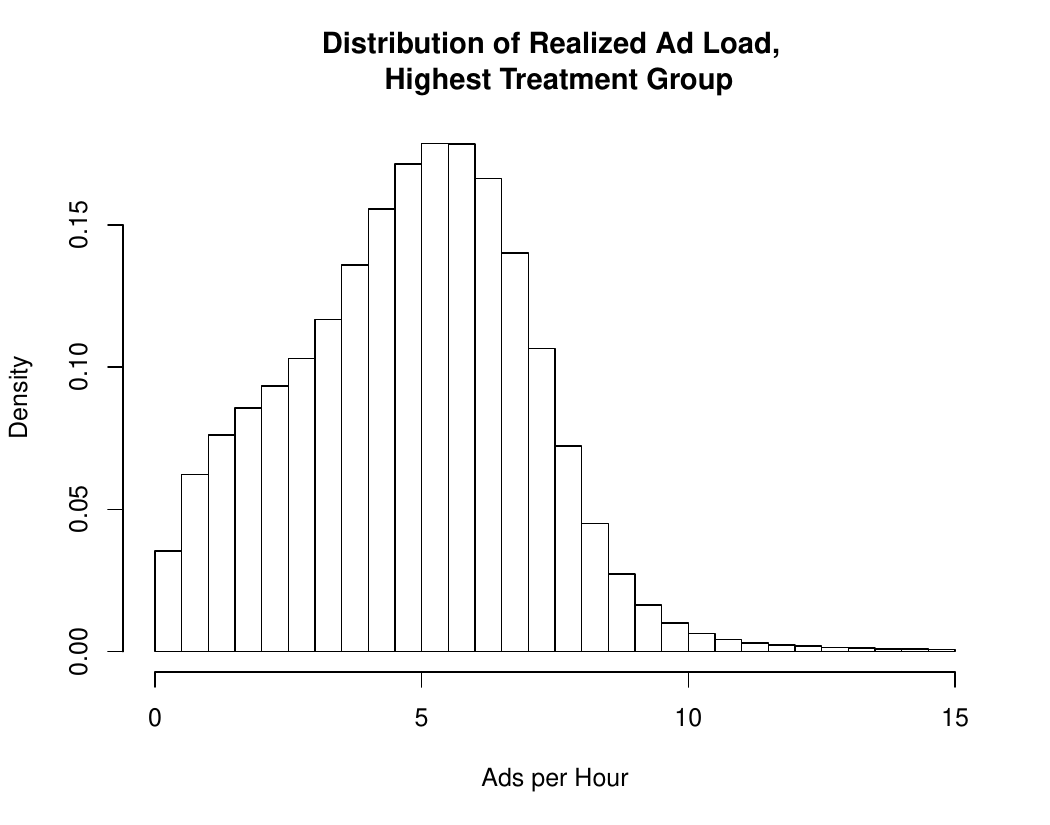}
\end{center}
\caption{Distribution of realized ad load within the highest intended treatment group. This histogram illustrates the range of actual ad loads experienced by users assigned to the same treatment condition.}
\label{fig:TG7-hist}
\end{figure}

While we will use the experiment to identify differences in listening behavior across treatment groups, it is interesting to note that there also exists considerable variation in ad load within each treatment group. Advertisers' demand varies, for example, by different ages, genders, and designated market areas (DMAs) of residence.\footnote{In section~\ref{sec:iv} we use this type of variation, i.e., persistent differences in advertisers' demand across age groups and DMAs, to construct instruments for measuring the effect of ad load on listening behavior. We then compare these estimates to those obtained from the experimental variation. The results show that these IVs yield severely biased estimates.} Figure \ref{fig:TG7-dmadiff} illustrates the variation in realized ad load across six randomly selected cities within the highest treatment group. The overall distribution of realized treatment within this treatment group can be seen in Figure \ref{fig:TG7-hist}. One might initially think that these persistent differences across DMAs could serve as a useful source of variation or a quasi-experiment for identifying the effect of changing ad load. However, as we demonstrate in Section \ref{sec:iv}, relying on this type of variation leads to severely biased estimates.

\section{Measuring the Sensitivity of Listeners to Advertising}
\label{sec:measurement}

In this section, we analyze how consumers respond to experimental changes in ad load using multiple outcomes: total hours listened, number of active days, and platform retention. Our overall measure of engagement is total hours (combining both ad-supported and subscription listening) as it provides a comprehensive metric that is well-defined regardless of a user's subscription status. For all outcomes, we normalize the metrics relative to the control group mean (multiplied by 100) to maintain confidentiality.\footnote{The normalization by control group mean has minimal impact on our standard errors. As shown in Online Appendix C, the adjusted standard errors change by less than 1\% of our estimates' magnitude due to our large control group of 18 million users.}

Our analysis proceeds in four stages. First, we document the aggregate impact of ad load variation on listening behavior and platform engagement. Second, we use instrumental variable regressions leveraging the experimental variation to estimate the marginal effect of increased ad load. Third, we compare the short-run and long-run effects of ad load changes. Finally, we decompose the treatment effect to understand the specific margins along which listeners adjust their consumption: hours per active day, days per active listener, and the probability of remaining active on the platform.

\subsection{Overall Impact on Platform Engagement}

\begin{figure}[!htbp]
\begin{center}
\includegraphics[scale = 0.45]{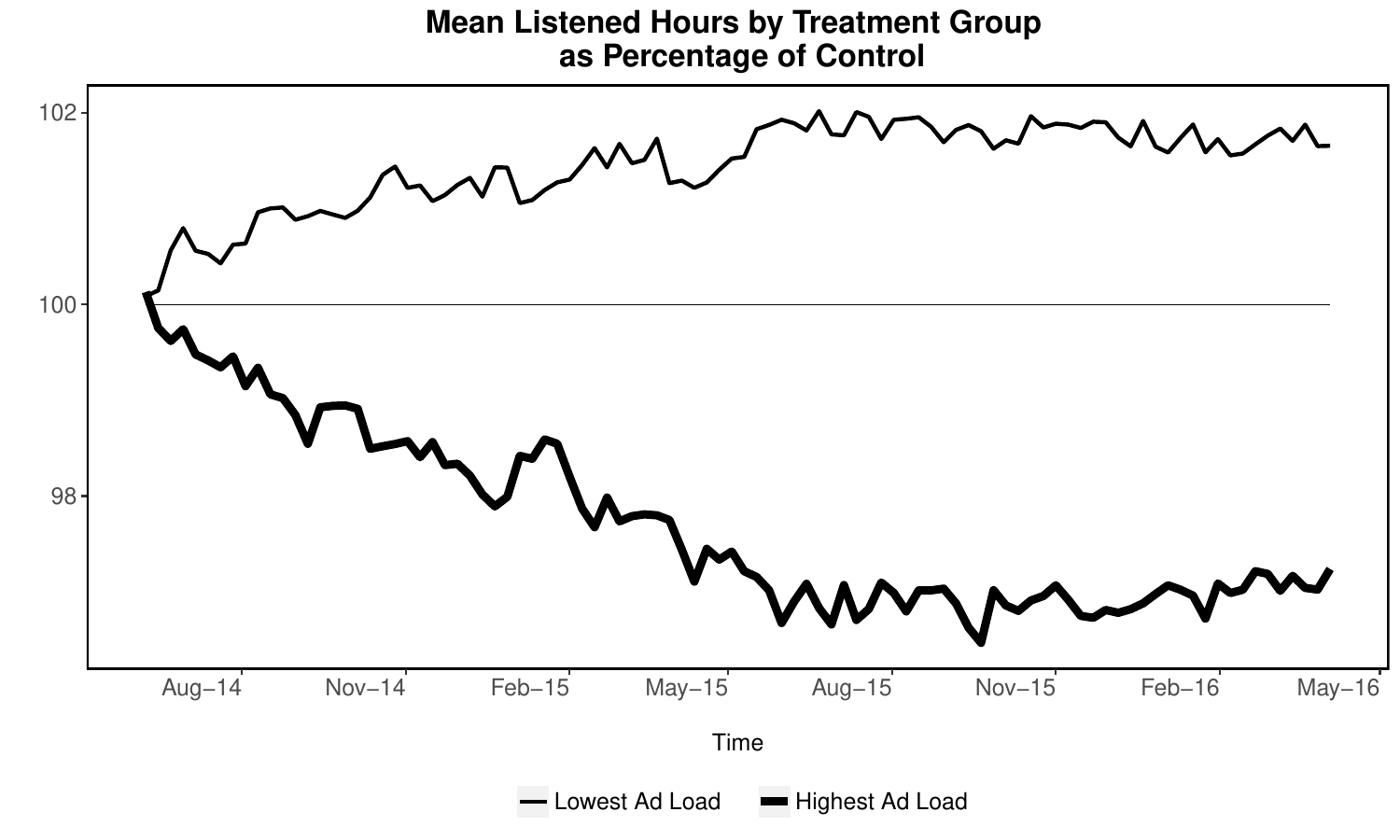}
\end{center}
\caption{ Mean weekly total hours listened by highest (6x2) and lowest (3x1) treatment groups over the course of the experiment, normalized relative to the control group.}
\label{fig:hours-intertemp}
\end{figure}

\begin{figure}[!htbp]
\begin{center}
\includegraphics[scale = 0.45]{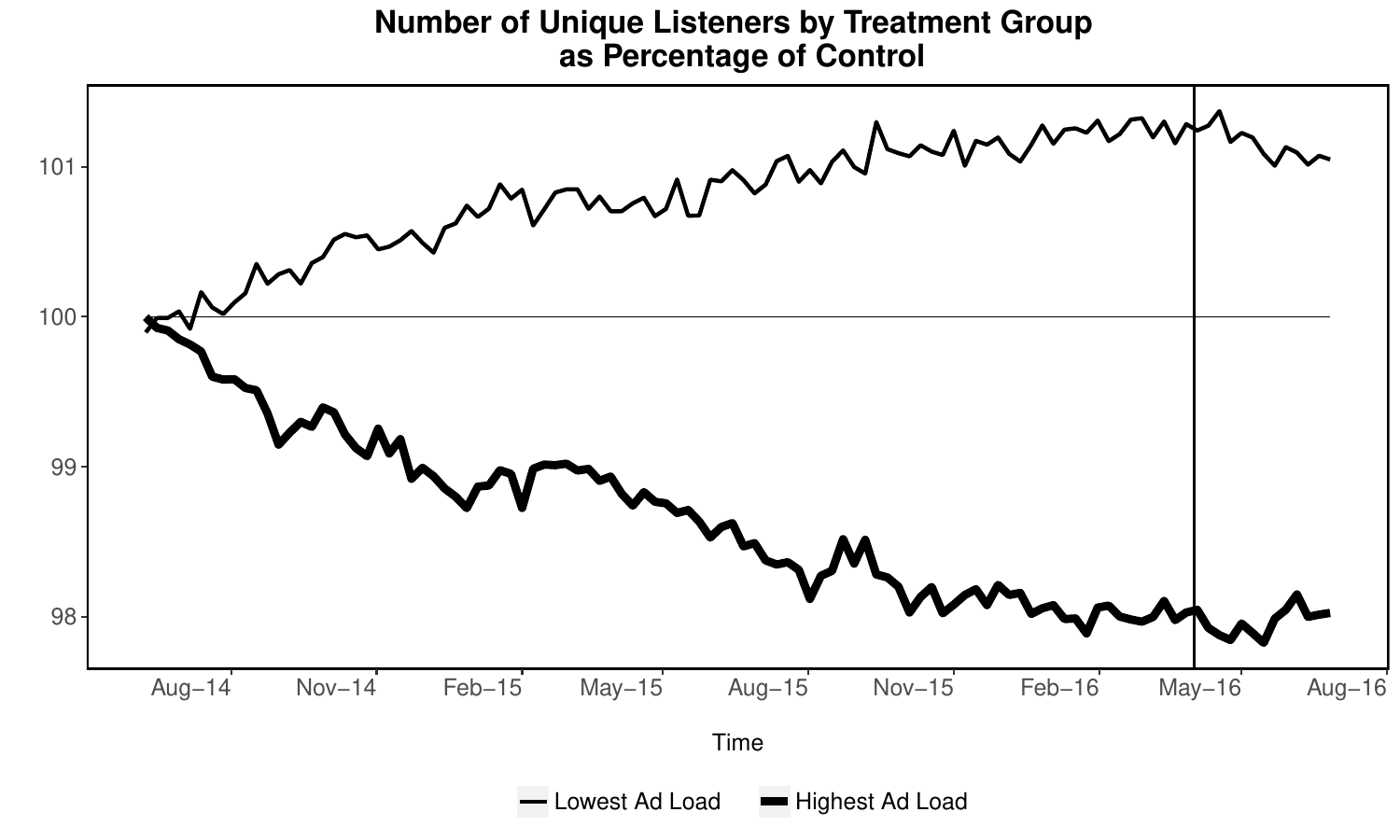}
\end{center}
\caption{Mean weekly unique listeners for highest (6x2) and lowest (3x1) treatment groups over the experiment duration, normalized relative to the control group.}
\label{fig:num-unique-intertemp}
\end{figure}

Figure \ref{fig:hours-intertemp} plots the number of hours listened each week in the highest and lowest ad-load treatments, each relative to the control group. Total hours listened diverges relatively gradually from that of the control group, with the highest ad-load treatment group gradually listening fewer and fewer hours relative to the control, while the lowest ad-load group gradually listens more and more hours. Figure \ref{fig:num-unique-intertemp} shows that this gradual change also holds true for the number of listeners actively listening to Pandora each week. By the end of the experiment, the highest treatment group has 2\% fewer weekly active listeners than the control, while the lowest treatment group has 1\% more listeners than the control. Most importantly, we see in both graphs of the weekly treatment effect how crucial it is that we ran the experiment for over a year. In both graphs, we observe that the treatment effect grows over the course of an entire year, stabilizing for the most part only after 12-15 months of treatment. Table \ref{table:lastmon-stats} shows that the treatment assignment impacted total hours and active days in a manner consistent with theory (i.e., users who were exposed to a higher ad load listened for fewer hours relative to the control, while those who were exposed to a lower ad load listened for more hours).

\begin{table}[ht]
\centering
\caption{Summary of realized total hours and days active in the final month (normalized relative to control). For data confidentiality, outcomes are normalized relative to their value in the control condition, which is why the control group always shows 100 for normalized variables.}
\label{table:lastmon-stats}
\resizebox{\textwidth}{!}{%
\begin{tabular}{lcccccccccc}
  \\ 
 \hline
 & 3x1 & 3x1.5 & 3x2 & 4x1 & 4x1.5 & 4x2 & 6x1 & 6x1.5 & 6x2 & Control \\ 
  \hline
  \hline
Total Hours (norm.) & 101.761 & 100.653 & 99.819 & 101.126 & 99.766 & 98.212 & 99.947 & 98.573 & 97.149 & 100.000 \\ 
  & (0.244) & (0.231) & (0.231) & (0.227) & (0.230) & (0.236) & (0.250) & (0.227) & (0.227) & (0.084) \\ 
  Days Active (norm.) & 101.716 & 100.727 & 99.685 & 101.171 & 99.709 & 98.689 & 100.201 & 98.827 & 97.365 & 100.000 \\ 
     & (0.135) & (0.134) & (0.133) & (0.134) & (0.133) & (0.133) & (0.134) & (0.133) & (0.132) & (0.042) \\ 
 Sample Size & 1,809,261 & 1,810,630 & 1,810,172 & 1,810,526 & 1,813,548 & 1,812,174 & 1,809,930 & 1,810,193 & 1,807,270 & 18,097,258 \\ 
   \hline
\end{tabular}%
}
\end{table}

\begin{figure}[!htbp]
\begin{center}
\includegraphics[width = 0.75\textwidth]{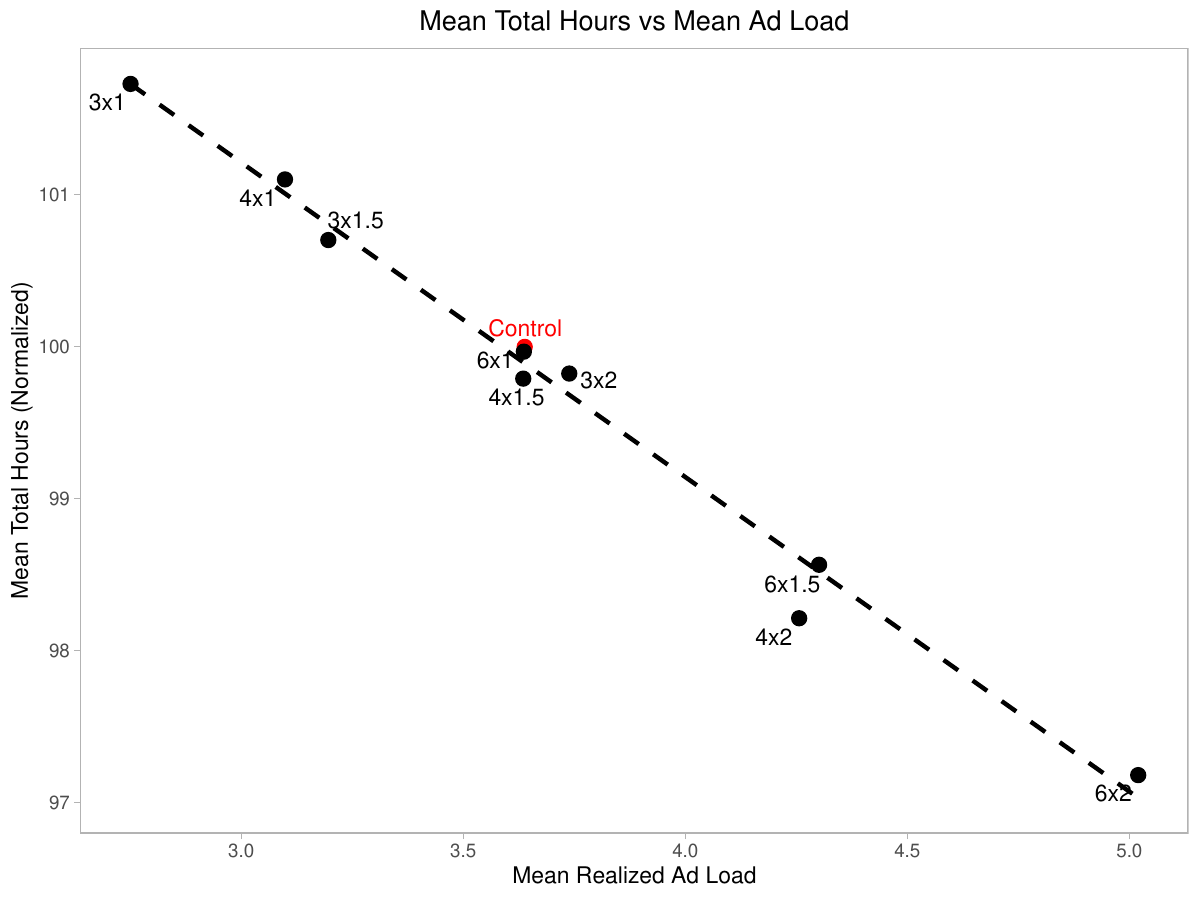}
\end{center}
\caption{Impact of ad load on final month hours per listener, by treatment group. Each point represents a treatment group, with the x-axis showing ads per hour and the y-axis showing normalized listening hours (relative to control).}
\label{fig:hours-overall}
\end{figure}

The experimental design allows us to measure the changes in listener behavior due to changes in the number of audio ads they receive. Figure \ref{fig:hours-overall} shows the estimated demand curve for Pandora listening in the final month of the experiment, as a function of the number of ads per hour received during the previous 21 months of treatment. Each of the nine treatment groups and the control group are plotted as a single point. We can see that this demand curve is strikingly linear. Since none of the points deviates very much from the best-fit line, we infer that the number of pods per hour versus the number of ads per pod have little influence on listening behavior, once we have modeled the effect of their product (ads per hour). We will return to this point below with an explicit model in section \ref{sect:interruptions}.

\subsection{Estimating Average Treatment Effect of Increasing Ad Load}

The best-fit line displayed in Figure \ref{fig:hours-overall} is the result of two-stage-least-squares (2SLS) estimation.  We use 2SLS instead of ordinary least squares (OLS) because we don't have complete control over treatment, for the variety of reasons discussed above, so that the realized ad load (number of ads per hour, aka ``price'') differs from what the experiment set as the intended maximum ad load. Here, the first stage regresses the realized ad load on the treatment dummy variables, which simply estimates the mean realized ad load in each treatment group, as illustrated by the horizontal coordinates in Figure \ref{fig:hours-overall}. The main equation (the second stage) regresses the outcome variable (hours listened in the final month) on the realized ad load, where the first stage causes us to use the mean ad load for each group (rather than the ad load for each individual) as the regressor.  The first stage guarantees that we exploit only the experimental differences between groups to identify the causal effects, removing all within-group variation that might yield spurious correlation - for example, urban listeners might both receive more ads and listen fewer hours, on average, than rural listeners, even though this correlation is not at all causal.  In addition to hours listened, we also consider the effects of treatment on several other outcome measures from the final month of the experiment.

\begin{model}[Simple IV -- Using Experimental Variation]
Let $y_i$ denote total listening hours or days for listener $i$ in the final month. The regression model we run is given by:
\begin{align}\label{eq:exp_iv}
y_i & = \beta_0 + \widehat{\left(\frac{ads}{hours}\right)}_i \cdot \beta_1 + \varepsilon_i, \\
\left(\frac{ads}{hours}\right)_i & = TG'_i \cdot \gamma + \eta_i, \nonumber
\end{align}
where $TG_i$ is a vector of treatment group indicators corresponding to each treatment condition.
\end{model}

\begin{table}[!htbp] \centering 
\caption{Using experimental variation to measure the effect of number of audio ads per hour on outcomes.}
  \label{table:simple-iv-wquad} 
\begin{tabular}{@{\extracolsep{5pt}}lcc} 
   \tabularnewline \midrule \midrule
   Dependent Variables: & Total Hours    & Active Days\\  
                  & (1)            & (2)\\  
   \midrule
   Ad Per Hour     & -2.082$^{***}$ & -1.911$^{***}$\\   
                        & (0.1157)       & (0.0662)\\   
   Constant             & 107.6$^{***}$  & 106.9$^{***}$\\   
                        & (0.4307)       & (0.2456)\\   
   \midrule
   Observations         & 34,390,962     & 34,390,962\\  
   F-test (1st stage) & 157,839.4      & 157,839.4\\  
   \midrule \midrule
   \multicolumn{3}{l}{\emph{Clustered (listener level) standard-errors in parentheses}}\\
   \multicolumn{3}{l}{\emph{Signif. Codes: ***: 0.01, **: 0.05, *: 0.1}}\\
\end{tabular} 
\end{table}

The 2SLS parameter estimates and standard errors are given in Table \ref{table:simple-iv-wquad}; we use heteroskedasticity-robust (White) standard errors when reporting all regression results in this paper. As shown in Figure \ref{fig:hours-overall}, the number of hours listened is a linearly decreasing function of the number of ads per hour during the treatment period.\footnote{The linearity was quite clear in Figure \ref{fig:hours-overall}, but we also checked the specification by adding a quadratic ad-load term or a logarithmic ad-load term to the linear term in the regression.  Unsurprisingly, both specifications produced small and statistically insignificant coefficients on the nonlinear terms.} The second column of Table \ref{table:simple-iv-wquad} shows that the number of days listened is similarly decreasing in the ad load (or price of listening). The coefficients show us that one additional ad per hour results in mean listening time decreasing by $2.082\%  \pm 0.227\%$, and the number of active listening days decreasing by $1.911\%\pm 0.130\%$.  We note that the large size of our experiment achieves considerable precision: the width of each 95\% confidence interval is less than one-fifth the size of the point estimate.

\subsection{Long-run versus Short-run Effects}

\begin{figure}[!htbp]
\begin{center}
\includegraphics[scale = 0.55]{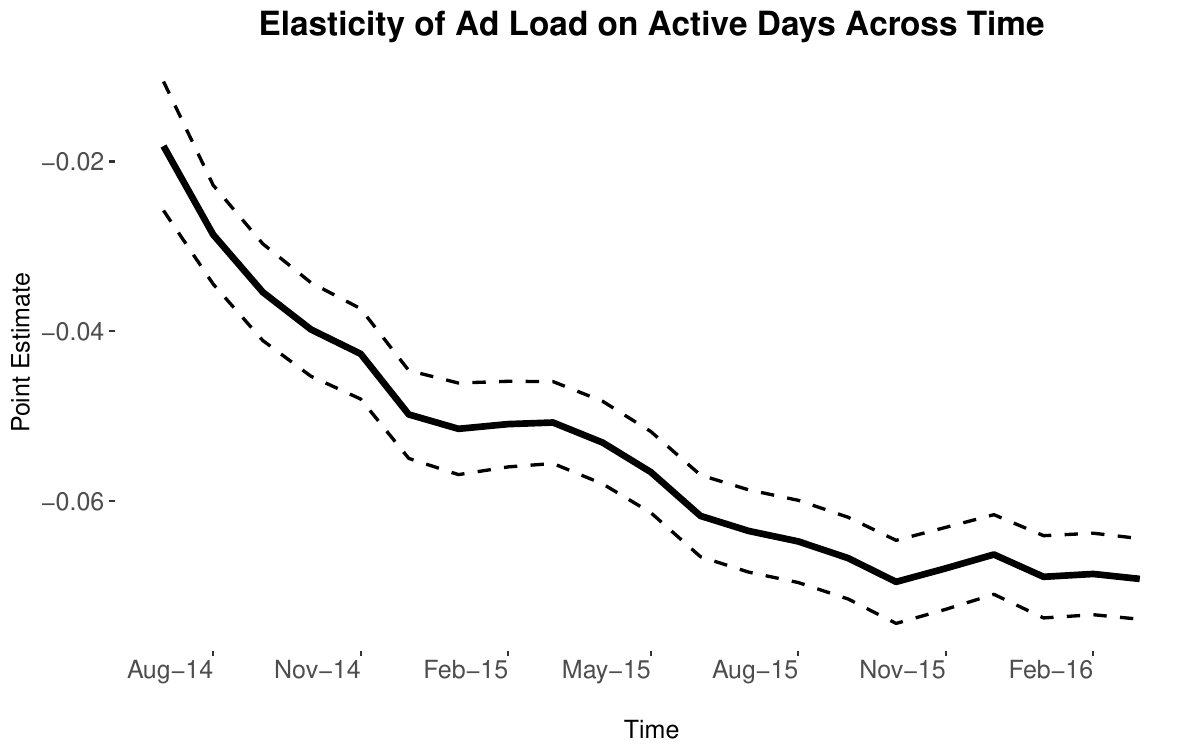}
\includegraphics[scale = 0.55]{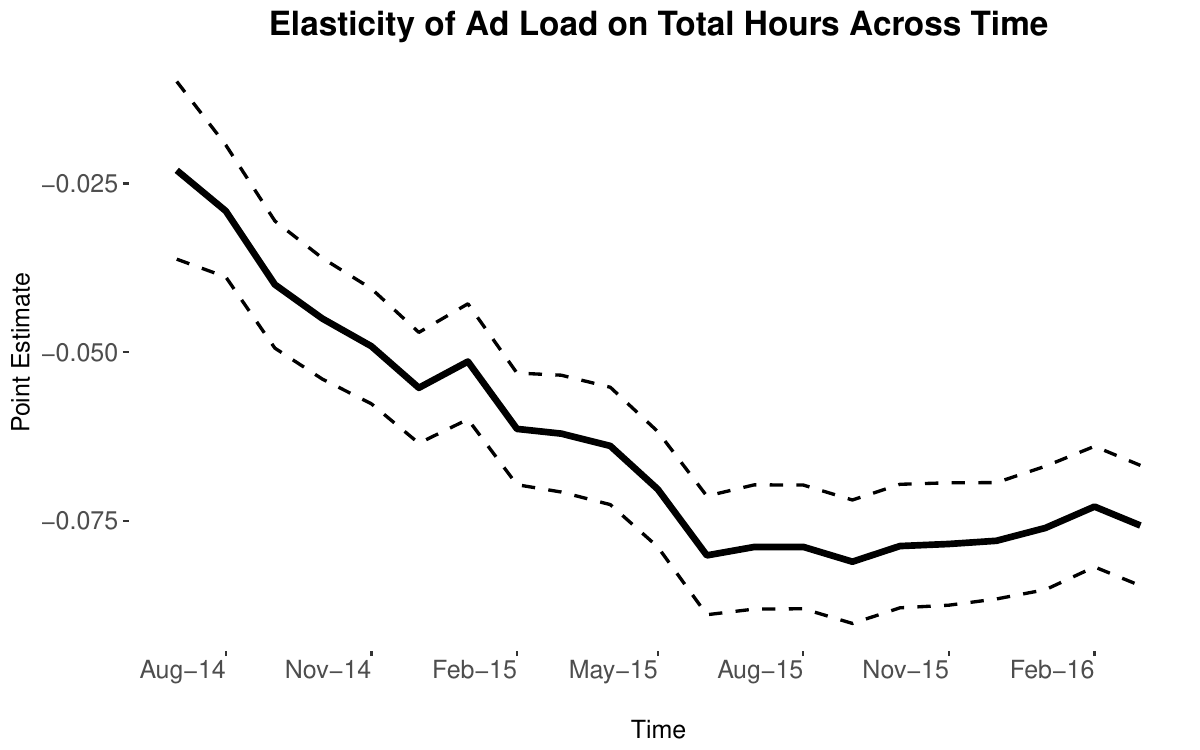}
\end{center}
\caption{Short- versus long-run elasticity of ad load on days active and total hours. The graphs show how the estimated elasticities change over the course of the experiment, demonstrating the importance of long-term measurement.}
\label{fig:temporal_estimates}
\end{figure}

How much does it matter that we conducted a long-run rather than a short-run experiment? To see how our estimates change with longer exposure to the treatment, we ran a 2SLS regression for each month of the experiment as if that month were the final one.\footnote{Since the experiment had continuous enrollment, the sample at each time point $t$ includes all users who had any ad-supported consumption during the experiment up until time $t$. } This time, instead of just a slope coefficient, we present the elasticity of demand with respect to changes in ad load. Figure \ref{fig:temporal_estimates} shows the estimated elasticity across time, with the solid line tracing the point estimates and the dotted lines tracing pointwise 95\% confidence intervals. For each month, we calculate the elasticity as the percentage change in consumption/activity divided by the percentage change in ad load. The slope in Figure~\ref{fig:hours-overall} is essentially the percentage change in consumption because of a unit increase in ad load. Note that a unit increase in ad load is a $1/3.622 = 27.6\%$ increase in ad load relative to the control condition. The elasticity is given by the slope divided by the percentage change in ad load of the control group from the beginning of the experiment to the end of that month. The estimated effects of a 1\% increase in ad load, on hours and days active, respectively, start out at around -0.02\% and -0.025\%, slowly increasing to effects of -0.070\% and -0.076\%. Had we run an experiment for just a month or two, we could have underestimated the true long-run effects by a factor of 3.

\subsection{Decomposing the Treatment Effect}
\label{sec:decompose}
Does this decrease in total listening come from shorter sessions of listening, or from a lower probability of listening at all?  To answer this question, Table \ref{table:iv-hours-tab} breaks the decrease in total hours down into three components: the number of hours listened per active day, the number of active days listened per monthly active listener, and the probability of being an active listener at all in the final month of the experiment. We have normalized each of these three variables so that the control group mean equals 100, so each of these treatment effects can be interpreted as a percentage difference from control. We find the percentage decrease in hours per active day to be approximately  0.4\%, the percentage decrease in days per active listener to be 0.95\%, and the percentage decrease in the probability of being an active listener in the final month to be 0.92\%. These three numbers sum to 2.27\%, which is approximately equal to the 2.08\% percentage decline we already calculated for total hours listened.\footnote{These three outcomes - hours per active day, days per active listener, and the probability of a listener being active - can be multiplied together to produce the total hours listened. Therefore, the sum of their percentage changes should equal the percentage change for the total hours. This arises from the fact that, for small changes, the percent change in a product is approximately the sum of the percentage changes: $$\frac{(X+\partial X)(Y + \partial Y) - XY}{XY} = \frac{\partial X}{X} + \frac{\partial Y}{Y} + \frac{\partial XY}{XY} \approx \frac{\partial X}{X} + \frac{\partial Y}{Y} $$The small disagreement between our two reported numbers likely results from the fact that the changes in ad load are not small ones.}

\begin{table}[htb]
\centering
\caption{Effect of audio ads per hour on listening outcomes, estimated using experimental variation (IV). The analysis focuses on four margins: total hours, activity, active days (conditional on activity), and hours per active day during the final month. This specification uses one observation per user and includes listeners from all experimental conditions.}
\label{table:iv-hours-tab}
\begin{tabular}{@{\extracolsep{5pt}}l*{6}{p{2cm}}}
\\[-1.8ex]\hline
\hline \\[-1.8ex]
Dependent Variables: & Total Hours & Active & Days/Active Listener & Hours/Active Day\\
& (1) & (2) & (3) & (4)\\
\midrule
Ad Per Hour & -2.082$^{***}$ & -0.9152$^{***}$ & -0.9545$^{***}$ & -0.4033$^{***}$\\
& (0.1157) & (0.0443) & (0.0475) & (0.0613)\\
Constant & 107.6$^{***}$ & 103.3$^{***}$ & 103.4$^{***}$ & 101.5$^{***}$\\
& (0.4307) & (0.1641) & (0.1703) & (0.2203)\\
\midrule
Observations & 34,390,962 & 34,390,962 & 14,109,882 & 14,109,882\\
F-test (1st stage) & 157,839.4      & 157,839.4       & 177,348.1            & 177,348.1\\  
\midrule \midrule
\multicolumn{5}{l}{\emph{Clustered (listener level) standard-errors in parentheses}}\\
\multicolumn{5}{l}{\emph{Signif. Codes: ***: 0.01, **: 0.05, *: 0.1}}\\
\end{tabular}
\end{table}

This analysis shows that the majority of the decline in listening hours is likely due to a decrease in the number of active days per listener and the likelihood of being an active listener at all. Approximately 18\% of the decline in the hours in the final month is due to a decline in the hours per active day, 42\% is due to a decline in the days per active listener, and 40\% is due to a decline in the number of listeners active at all on Pandora in the final month.

However, this analysis faces two potential selection issues. First, increased ad load leads to churn off the platform, which means the set of users who remain active may systematically differ in their listening patterns. Second, more ad-sensitive users may be more likely to churn early, leaving a selected sample of less sensitive users, which could lead to underestimation of the effect on consumption conditional on activity. In Online Appendix D, we provide additional evidence by combining the panel structure of our data with experimental variation to address these selection concerns. While selection issues cannot be fully ruled out, the analyses support our finding that the primary margin of adjustment is through reduced frequency of listening rather than shortened session lengths.

The results from the panel specification (Table \ref{table:breakdown_IV_panel}) are remarkably similar to those from the simple IV specification (Table \ref{table:iv-hours-tab}), despite using a different sample and methodology. The panel estimates show that one additional ad per hour reduces mean listening time by 2.157\% (SE: 0.182\%), compared to 2.082\% (SE: 0.116\%) in the simple IV specification. Similarly, the effect on active days conditional on activity (-0.958\% vs -0.955\%), and hours per active day (-0.431\% vs -0.403\%) are nearly identical across both specifications.

These findings allow us to better understand how users adjust their listening behavior in response to increased ad load. Conditional on users starting a session, consumption appears relatively inelastic, possibly because listening duration is often determined by external factors like commute time. However, while we observe adjustments across all margins, the primary effect stems from users being less likely to initiate listening sessions and remain active. Over time, users appear to have both learned about the change in service quality and adjusted their listening habits accordingly, with the strongest response being fewer initiated sessions. While all margins across both panel and cross-sectional IV specifications show statistically significant reductions, the majority of the effect involves fewer listening sessions rather than reduced hours per session. This analysis contributes to our understanding of how increased ad load differentially affects various aspects of user engagement.

\section{Causality from Experimental Versus Observational Data} 
\label{sec:observational}

How valuable is the variation generated by this experiment? Since it can be difficult to convince decision-makers to run experiments on key economic decisions, and it consumes engineering resources to implement such an experiment properly, could we have done just as well by using observational data? To investigate this question, we reran our analysis using only the 18 million listeners in the control group, since they were untouched by the experiment in the sense that they received the default ad load. We employed three different methods with observational data: (a) OLS and OLS with high-dimensional controls and FEs, (b) IV regressions based on instruments that reflect advertisers' differential interest across different age groups and DMAs, and (c) panel regressions.

Below, we will discuss and compare the performance of these methods using the experimental variation to highlight the necessity and value of running controlled experiments for obtaining accurate and reliable estimates of the impact of ad load on listening behavior.

\subsection{Cross-sectional regressions}
\label{sec:OLS_FE}

We start by regressing listening hours and active days in the final month on ad load. We consider the following specification:
\begin{model}[Cross-sectional OLS]
Let $y_{i}$ denote the outcome of interest for listener $i$ in the final month of the experiment:
\begin{align}\label{eq:OLS}
y_i = \beta_0 + \frac{ads}{hours_i} \beta_1 + \epsilon_i,
\end{align}
The parameters are defined similarly to~\eqref{eq:exp_iv}; however, here we do not instrument for the ad load using the experimental condition. Instead, we only focus on users in the control group and rely on natural variation in the data. \end{model}

We estimate model~\eqref{eq:OLS} using total hours and active days as the dependent variables. The results are reported in columns (1) and (4) of Table~\ref{tab:OLS_FE_IV}. Comparing the results with those estimated using experimental variation in Table~\ref{table:simple-iv-wquad}, the estimate for total hours is magnified, whereas the estimate for active days is shrunk toward zero. One obvious criticism of a cross-sectional regression is that it suffers from omitted variable bias: unobservable listener-level heterogeneity may produce a non-causal correlation between the ad load and the dependent variable. One common approach to addressing this issue is to use controls and high-dimensional fixed effects to absorb confounding variables that might affect the analysis. We consider the following model:
\begin{model}[Cross-sectional OLS with Fixed Effects and Controls]
Let $y_{i}$ denote the outcome of interest for listener $i$ in the final month of the experiment:
\begin{align}\label{eq:OLS_FE}
y_i = \beta_0 + \frac{ads}{hours_i} \cdot \beta_1 + \boldsymbol{X}_i \cdot \beta_2 + \delta_{age_i} + \delta_{zip_i}  + \epsilon_i,
\end{align}
The parameters are defined similarly to~\eqref{eq:OLS}; however, we now include a number of pre-treatment features $\boldsymbol{X}i$ and high-dimensional fixed effects $\delta_{age_i}$ and $\delta_{zip_i}$. The control variables include hours, days active, number of ad pods, and number of ads served in the May 2014 (pre-treatment) period. The fixed effects include user age and zip code. \end{model}

The results from specification~\eqref{eq:OLS_FE} are presented in columns (2) and (5) of Table~\ref{tab:OLS_FE_IV}. Although specification~\eqref{eq:OLS_FE} controls for cross-sectional differences across a number of behavioral variables, such as consumption, and uses fixed effects to absorb differences across zip codes and age groups, the results end up becoming even more biased than those in columns (1) and (4) of Table~\ref{tab:OLS_FE_IV} (simple OLS) compared to the estimates from the experimental variation in Table~\ref{table:simple-iv-wquad}.

This finding is concerning because applied economists often use high-dimensional fixed effects and cross-sectional features to reduce bias, and one would typically expect the estimates to move towards the ``true'' causal coefficients. However, control variables and high-dimensional fixed effects may absorb some endogenous variation while also absorbing quasi-experimental variation (good variation) in ad load across different age groups and DMAs that might help identify the causal effect of ad load. Unfortunately, this exercise demonstrates that with observational data, it is not possible to determine whether the inclusion of high-dimensional fixed effects and controls reduces or increases the bias. This outcome emphasizes the importance of having experimental variation in order to accurately estimate causal effects.

\begin{table}
\centering
\caption{Using cross-sectional OLS, OLS with high-dimensional controls and fixed effects, and IV regressions to measure the effect of ad load on total hours and active days.}
\label{tab:OLS_FE_IV}
\resizebox{0.9\textwidth}{!}{%
\begin{tabular}{lcccccc}
   \tabularnewline \midrule \midrule
   Dependent Variables: & \multicolumn{3}{c}{Total Hours} & \multicolumn{3}{c}{Active Days}\\
   Model:                 & (1)            & (2)            & (3)            & (4)            & (5)            & (6)\\  
   \midrule
   Ad Per Hour       & -3.462$^{***}$ & -6.254$^{***}$ & -11.51$^{***}$ & -1.097$^{***}$ & -4.217$^{***}$ & -8.854$^{***}$\\   
                          & (0.0429)       & (0.3183)       & (0.8353)       & (0.0130)       & (0.0116)       & (0.5135)\\   
   Constant               & 112.6$^{***}$  &                &                & 104.0$^{***}$  &                &   \\   
                          & (0.2181)       &                &                & (0.0643)       &                &   \\   
   \midrule
   Pre-treatment controls &              & X            &              &              & X            & \\  
   \midrule
   Zip code FE               &                & X            & X            &                & X            & X\\  
   Listener age FE             &                & X            & X            &                & X            & X\\  
   \midrule
   Observations           & 18,097,258     & 18,097,258     & 18,097,258     & 18,097,258     & 18,097,258     & 18,097,258\\  
   F-test (1st stage) &                &                & 27,135.0       &                &                & 27,135.0\\ 
   \midrule \midrule
   \multicolumn{7}{l}{\emph{Clustered (listener level) standard-errors in parentheses}}\\
   \multicolumn{7}{l}{\emph{Signif. Codes: ***: 0.01, **: 0.05, *: 0.1}}\\
\end{tabular}}
\end{table}

\subsection{Instrumental Variable Regressions without Experimental Variation}
\label{sec:iv}
There are two major concerns when using OLS to estimate the causal effect of long-run changes in ad load on listening behavior. First, individual-level differences in behavior can lead to confounding, and controls or fixed effects may not sufficiently account for these differences. For instance, ad frequency caps make it harder to fill ad slots for users who listen for longer hours, causing ad load to drop for these users and inflating the effect of ad load on listening hours. Second, OLS may capture temporary changes in ad load rather than its long-term effects. To measure the long-run effect of ad load changes in equilibrium, we need persistent differences in ad load across user groups that are otherwise similar.

One approach is to use instruments that shift ad load for a population or reflect persistent aggregate-level differences in advertisers' preferences for these groups but are not correlated with individual users' listening behavior. As discussed, Pandora audio ads are sold via forward contracts with attributes like age group and DMA. Therefore, users from different age groups and DMAs might receive persistently different levels of ad load, and this variation can be used to identify the long-run effect of ad load changes on listening behavior.

\begin{figure}[htbp]
    \centering
    \begin{subfigure}[b]{0.45\textwidth}
        \centering
        \includegraphics[width=\textwidth]{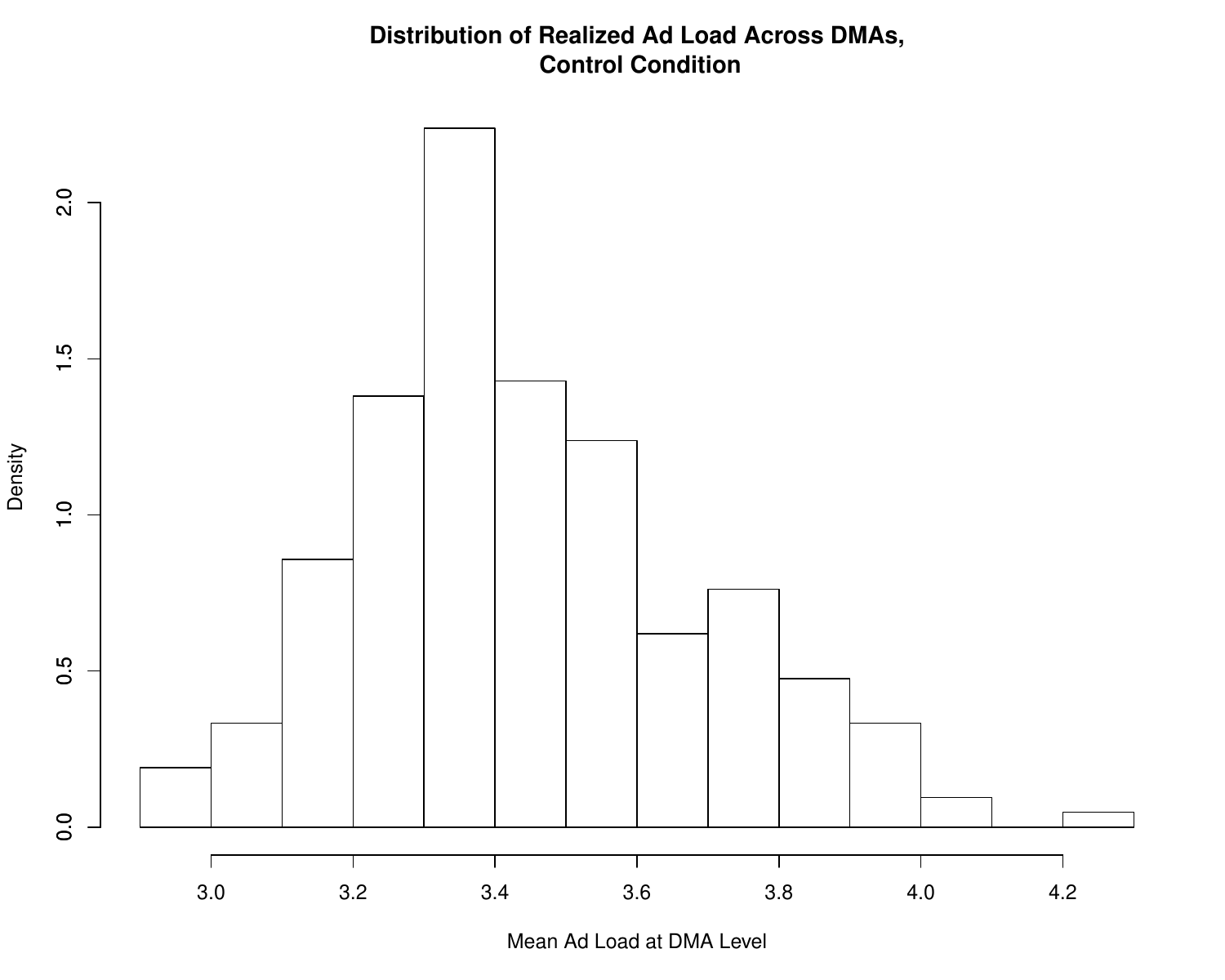}
    \end{subfigure}
    \hfill
    \begin{subfigure}[b]{0.45\textwidth}
        \centering
        \includegraphics[width=\textwidth]{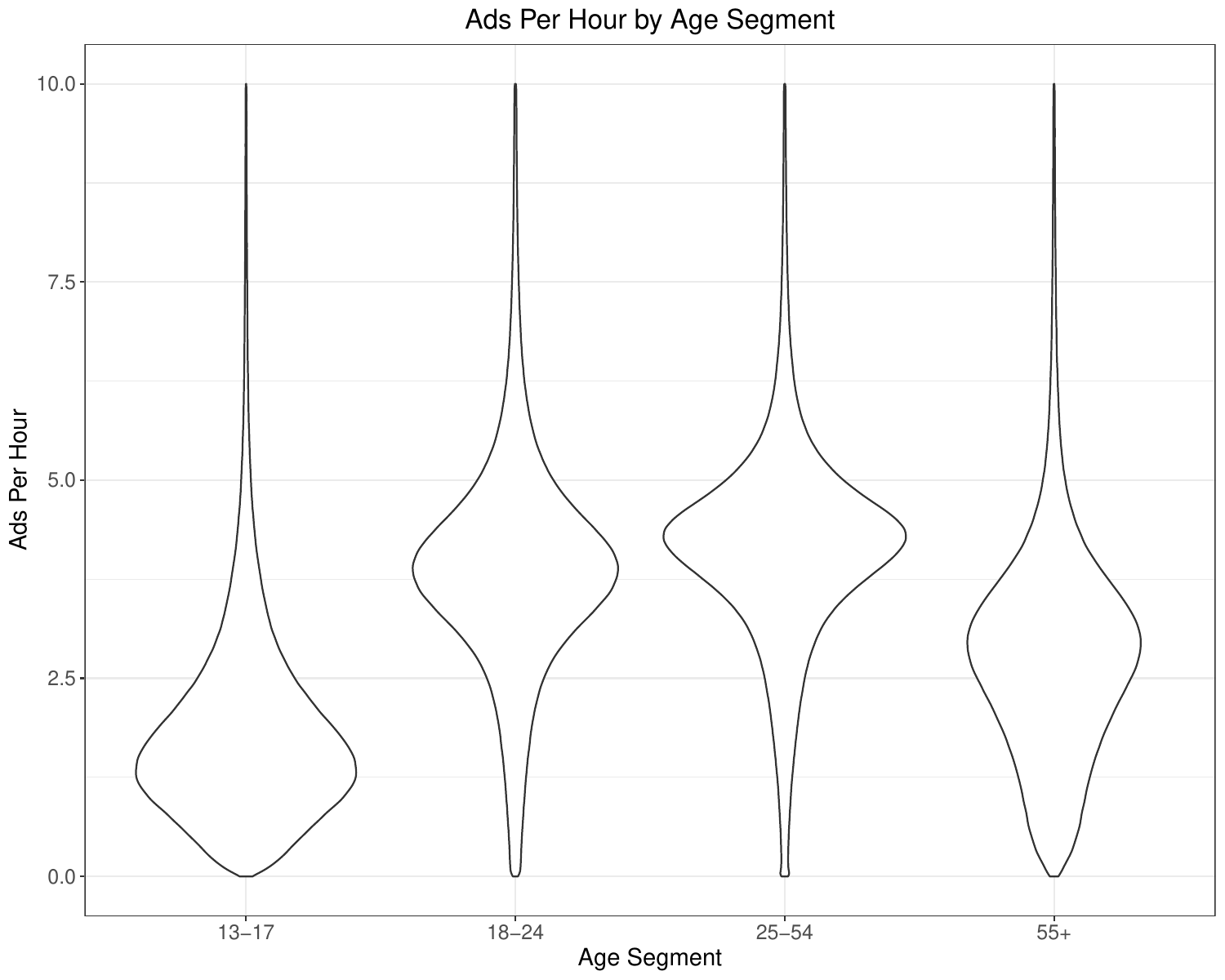}
    \end{subfigure}
\caption{Variation in realized ad load under the control condition across different DMAs (left panel) and different age groups (right panel).}
\label{fig:variation_ad_load_dma_age}
\end{figure}

Figure~\ref{fig:variation_ad_load_dma_age} shows the average ad load across different DMAs and age groups under the control condition, indicating significant heterogeneity likely due to advertisers' differential interest. While the average ad load in a DMA reflects advertisers' demand, listening habits might still differ across DMAs, potentially violating the exclusion restrictions when using the average ad load in a DMA as an instrument. A similar argument applies to age groups; differences in ad load across age groups may also be correlated with listening habits.

One might argue that using the average realized ad load within an age group and a DMA as an instrument while controlling for geographic and age differences (fixed effects) could be a viable approach. The argument is that after controlling for age and geographic fixed effects, a user in the 25-54 age group in New York DMA would be comparable to a user in the same age group in Los Angeles DMA, and persistent differences in ad load across DMAs for these age groups would only affect users through the realized ad load. However, even after accounting for cross-sectional differences across different DMAs, similar age groups in different DMAs might still have different listening habits due to interactive effects between geographic and demographic factors, potentially violating exclusion restrictions. Nevertheless, such an instrumental variable approach that leverages persistent cross-sectional differences in exposure to treatment across users has been used in other contexts. For instance, channel positions across zip codes for exposure to conservative media \citep{martin2017bias}, mortality rates among settlers as an IV for institutions \citep{acemoglu2001colonial}, and group populations as instruments for radio stations catering to different demographics \citep{waldfogel2003preference}.

To operationalize this, we consider the following model:
\begin{model}[IV Regressions without Experimental Variation]
Let $y_i$ denote total listening hours or days for listener $i$ in the final month. The regression model we run is given by:
\begin{align}\label{eq:iv_no_exp}
y_i & = \beta_0 + \widehat{\frac{ads}{hours}}_i \cdot \beta_1 + \delta_{age_i} + \delta_{zip_i} + \epsilon_i, \\
\frac{ads}{hours}_i & = \bar{\mathcal{A}}_{\{DMA_i,age_i\}} \cdot \gamma + \delta_{age_i} + \delta_{zip_i} + \eta_i. \nonumber
\end{align}

The parameters are defined similarly to~\eqref{eq:OLS_FE} and~\eqref{eq:exp_iv}; however, $\bar{\mathcal{A}}_{\{DMA_i,age_i\}}$ represents the average ad load for the DMA and age group to which user $i$ belongs.
\end{model}

We would like to emphasize again that the IV $\bar{\mathcal{A}}$ is cross-sectional and captures persistent differences in ad load across age groups in different DMAs. Consequently, we cannot control for any endogenous (pre-treatment) features, such as listening hours, because these were also ``treated'' by the same cross-sectional differences (IV). Appendix C of~\citet{acemoglu2001colonial} provides a discussion on this issue. Therefore, in this specification, we only control for zip code and age fixed effects.

The estimates and first-stage F-statistics from specification~\eqref{eq:iv_no_exp} are presented in columns (3) and (6) of Table~\ref{tab:OLS_FE_IV}. The first-stage F-statistics is 27,135, indicating that the instrument is very strong. However, the coefficients end up being even more inflated than those from the OLS and OLS with fixed effects (columns (1)-(2) and (4)-(5) of Table~\ref{tab:OLS_FE_IV}). This finding is concerning because researchers without access to experimental data might compare the results with fixed effects to the IV results and, seeing that both methods yield larger coefficients than OLS, incorrectly conclude that the OLS coefficients are biased downwards. However, the results from the experimental variation in Table~\ref{table:simple-iv-wquad} do not support this conclusion.

This discrepancy highlights the importance of having experimental variation to accurately estimate causal effects. In the absence of experimental data, researchers might draw incorrect conclusions about the direction and magnitude of bias in observational estimates. The inflated IV estimates in this case demonstrate that econometric techniques, such as instrumental variables and fixed effects, can produce misleading results when the underlying assumptions are violated or when there are complex interactions between the treatment and other factors.

These findings complement and extend the recent work by~\citet{moshary2021and} on the use of instrumental variables for studying the effect of commercial advertising. While they focused primarily on weak instrument concerns, finding that political advertising was a weak instrument for most product categories, our results highlight a different but equally important issue. In our case, we have a strong instrument (as evidenced by high first-stage F-statistics), yet it still leads to misleading results when compared to our experimental findings. This discrepancy reveals the subtle ways in which the exclusion restriction of IVs might be violated in practice. Our experimental results allow us to detect these violations, which would not be apparent from observational data alone. This underscores that even when standard diagnostic tests suggest an instrument is strong, the resulting estimates may still be biased or imprecise due to violations of the exclusion restriction. Our ability to compare these results to experimental findings provides a unique opportunity to validate (or in this case, challenge) the reliability of IV approaches in advertising research.

In the next section, we explore the use of panel variation to examine whether panel regression results are less biased compared to the observational methods employed thus far. By leveraging within-user variation in ad load over time, we aim to mitigate some of the confounding factors that plague cross-sectional estimates. However, as we will discuss, panel methods also have their limitations and may not fully address the endogeneity concerns in estimating the causal effect of ad load on listening behavior.

\subsection{Using Panel Variation}

Another valuable technique for performing causal inference with observational data is to use a panel regression with listener and time fixed effects. However, without a long-run experiment, we would not have known the relevant time period to use in our panel estimation. We might have chosen to look at variation from a single month, which clearly would not have produced the correct long-run causal effects. To give the panel estimator its best shot, we use what we have learned from the experiment and allow the panel estimator a 20-month time period between observations. We consider the following model:
\begin{model}[Panel Regression]
Let $y_{it}$ denote the outcome of interest for listener $i$ in month $t$, where $t$ is either the month prior to the start of the experiment or the final month. The regression model we run is given by:
\begin{align}\label{eq:panel}
y_{it} & = \delta_i + \tau_t + \frac{ads}{hours}_{it} \cdot \beta_1 + \varepsilon_i,
\end{align}
where $\delta_i$ and $\tau_t$ are listener and month fixed effects, respectively. For the final month, $\frac{ads}{hours}_{it}$ is calculated based on the experimentation period while for the month prior to the experiment, that ratio is calculated based on May of 2014. To ensure a balanced panel, we restrict the sample to listeners active in May of 2014. 
\end{model}

To ensure a valid comparison and consider the appropriate population, we need to address a key difference in our analyses. For the panel regression, we constrained the sample to listeners who were active in May 2014. This sample selection means that the results are not directly comparable to those in Table~\ref{table:simple-iv-wquad}, which includes a broader set of users. To create a more equivalent comparison, we re-estimate specification~\eqref{eq:exp_iv} by selecting only users who were active in May 2014 and relying on experimental variation. This approach measures the average treatment effect of an increase in ad load specifically on users who were active at the start of our study period, allowing for a more accurate comparison between our different analytical methods.

Table~\ref{table:panel} displays the estimates using panel variation on the control users who were active in May 2014 (columns (1)-(2)) and the estimates obtained by pooling active users in May 2014 across all experimental conditions by leveraging the experimental variation (columns (3)-(4)). We see from Table~\ref{table:panel} that the point estimate for active days in column (2) is much closer to that found in column (4), which is based on experimental variation, but it still overestimates the impact of ad load by more than the width of our 95\% confidence intervals. The panel point estimate for total hours still overestimates the effect by a factor of 3.

Similar to our discussion above, we suspect that one likely reason for the large bias in the effect of ad load on hours is because audio ad campaigns on Pandora have hourly and daily frequency caps, such as ``no listener should hear this ad campaign more than three times in a day.'' We often find that listeners who listen for many hours in a day end up exhausting the supply of ads targeting them, so after several hours, their ad load may diminish because Pandora has no more eligible ads left to play for them. Thus, a reverse-causality argument applies - listening more hours in a day causes ad load to diminish - and this provides a strong source of bias in our panel estimator for the effects of ad load on listening hours. The experimental estimate, of course, has no such difficulty because we exploit only the exogenous variation in making those estimates.

Our results suggest that, even after controlling for time-invariant listener heterogeneity, observational techniques still suffer from omitted-variable bias and even reverse causality, caused by unobservable terms that vary across individuals and time that correlate with ad load and listening behaviors. Without a long-run experiment, we would not have known the relevant timescale to consider to measure the long-run sensitivity to advertising, which is what matters for the platform's policy decisions.

\begin{table}[!htb] \centering 
\caption{Comparison of estimates from panel regression using control-condition users active in May 2014 (columns 1-2, two observations per listener) with estimates from instrumental variable regression using experimental variation across all conditions for users active in May 2014 (columns 3-4, one observation per listener).}
  \label{table:panel} 
\begin{tabular}{lcccc}
   \tabularnewline \midrule \midrule
   Dependent Variables: & Total Hours            & Active Days             & Total Hours             & Active Days\\  
   Model:               & (1)              & (2)              & (3)               & (4)\\  
   \midrule
   Ads Per Hour         & -6.62629$^{***}$ & -2.35709$^{***}$ & -2.33631$^{***}$  & -2.12157$^{***}$\\   
                        & (0.06267)        & (0.02499)        & (0.14204)         & (0.08149)\\   
   Constant             &                  &                  & 108.91244$^{***}$ & 108.16536$^{***}$\\   
                        &                  &                  & (0.55914)         & (0.31982)\\   
   \midrule
   Period               & X              & X              &                   & \\  
   Listener         & X              & X              &                   & \\  
   \midrule
   Observations         & 13,928,268       & 13,928,268       & 13,235,071        & 13,235,071\\  
   F-test (1st stage)   &                &                & 100,034.5      & 100,034.5\\     
   \midrule \midrule
   \multicolumn{5}{l}{\emph{Clustered (listener level) standard-errors in parentheses}}\\
   \multicolumn{5}{l}{\emph{Signif. Codes: ***: 0.01, **: 0.05, *: 0.1}}\\
\end{tabular}
\end{table}

\begingroup
\centering
\par\endgroup

\section{Business and Managerial Implications}
\label{sec:business}

Our results in section~\ref{sec:measurement}, and in particular Figure~\ref{fig:temporal_estimates}, reflect the ad load elasticity of demand in the short and long run. Our findings show that the long-run elasticity is three times larger than the short-run ad load elasticity of demand (see Figure~\ref{fig:temporal_estimates}). Furthermore, the estimates are smaller than those measured using observational methods. These results alone demonstrate that experiments are crucial, as stakeholders and decision-makers might select lower ad loads based on observational data. While these findings have significant business implications, in this section, we also show that experimental data can be valuable for making decisions in a few other key areas. Below, we first investigate the effect of frequency and length of commercial breaks and their implications for optimal ad scheduling. Then, we discuss how ad load can impact demand for Pandora Plus, the subscription-based product, and how using observational data in both cases can be misleading.

\subsection{The Effect of the Length and Frequency of Commercial Interruptions} \label{sect:interruptions}

How do consumers react to the number of commercial interruptions, as opposed to merely the number of ads? Recall that the experimental design separately varied two different components of the ad load: the number of pods per hour and the number of ads per pod. Earlier, we noted that the main impact on listening comes through their product, the total number of ads per hour. However, the distribution of ads across pods may also matter. For instance, \citet{yao2017tv} demonstrate in their study of TV viewing behavior that commercial breaks provide opportunities for viewers to search for alternatives. They model TV viewing as a sequential search process where viewers can explore other channels during these breaks.

\begin{figure}
    \centering
    \includegraphics[width = 0.8\textwidth]{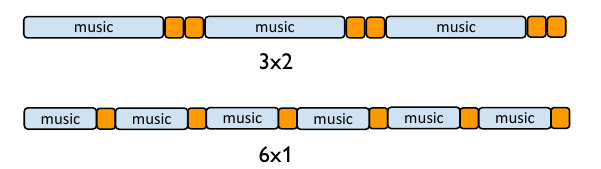}
\caption{Illustration of two different ad scheduling strategies: 3x2 (three pods with two ads each) and 6x1 (six pods with one ad each). This diagram shows how the same total ad load can be distributed differently across a listening hour.}
    \label{fig:pod_freq_len}
\end{figure}

To illustrate this in our context, consider two different treatment conditions: the 3x2 and the 6x1 conditions, as depicted in Figure~\ref{fig:pod_freq_len}. These conditions deliver a similar ad load but differ in ad scheduling. In the 6x1 condition, six pods (ad breaks) with one ad each are played per hour, whereas in the 3x2 condition, three pods containing two ads each are served to the listener. While the total number of ads is the same, the 6x1 condition creates more frequent but briefer interruptions to the listening experience. In contrast, the 3x2 condition results in fewer but more prolonged disruptions, which might increase the likelihood of listeners seeking alternatives during these extended breaks.

Intuitively, if only the overall ad load matters to the listener, we would expect that serving one ad pod of length 2 per hour would have a similar effect on consumption as two ad pods of length 1 per hour. However, if the number and length of search opportunities are important, as suggested by the framework in \citet{yao2017tv}, we might expect different effects. To measure the effect of the ad serving strategy, we count the number of pods of length 1 and 2 per hour and denote them by $\mathcal{P}_1$ and $\mathcal{P}_2$. Note that the overall ad load would be equal to $\mathcal{P}_1 + 2 \cdot \mathcal{P}_2$. To measure the impact of the number of commercial breaks versus the length of each commercial break, we consider the following model.

\begin{model}[Effect of Pod Length and Frequency]
Let $y_i$ denote total listening hours and days in the final month for listener $i$. The regression model we run is given by:
\begin{align}\label{eq:simp_iv_app}
y_i & = \beta_0 + \widehat{\mathcal{P}_{1,i}} \beta_1 + \widehat{\mathcal{P}_{2,i}} \beta_2  + \varepsilon_i, \\
\left(\frac{\text{pods of length 1}}{hours}\right)_i \rightarrow \mathcal{P}_{1,i}  &= TG'_i \gamma_1 + \eta_{1,i}, \nonumber \\
\left(\frac{\text{pods of length 2}}{hours}\right)_i \rightarrow \mathcal{P}_{2,i}  &= TG'_i \gamma_2 + \eta_{2,i}. \nonumber
\end{align}
\end{model}

As highlighted above, the experiment not only shifts the overall ad load but also alters the composition of pods (pods of length 1 versus pods of length 2). The instrumental variable regression model in~\eqref{eq:simp_iv_app} leverages this variation to measure the effect of pod length and frequency. The estimates from model~\eqref{eq:simp_iv_app} are displayed in Table~\ref{table:pod_len_freq}. The effect of pods of length 2 and 1 on total hours is -4.12 and -1.97, respectively. Similarly, the effect of an additional pod of length 2 and 1 on active days is equal to -3.65 and -1.87. As expected, the impact of pods of length 2 is approximately twice as large as pods of length 1. If listeners are indifferent to the ad scheduling method and only respond to the overall ad load rather than pod length or frequency, then $2\cdot \beta_1 \approx \beta_2$.

To formally test this hypothesis, we conduct a null hypothesis test that $2\cdot \beta_1 - \beta_2 = 0$ and report the results in the bottom section of Table~\ref{table:pod_len_freq}. While the results from both tests indicate a positive sign ($2\cdot \beta_1 - \beta_2 > 0$), we can only reject the null hypothesis for active days at $p < 0.05$. Therefore, We find weak evidence that listeners preferred 2 pods of length 1 to a pod of length 2. This suggests that listeners respond primarily to total ads per hour and might have a slight preference for shorter but more frequent pods rather than longer but less frequent breaks, as indicated by $2\cdot \beta_1 > \beta_2$. It is worth noting that had we found $2\cdot \beta_1 < \beta_2$, it would have implied that users prefer an ad pod of length 2 to two ad pods of length 1, suggesting a preference for fewer but longer ad pods. In Online Appendix A, we demonstrate that using observational data produces misleading results, and in some cases, even the sign for $\mathcal{P}_{2,i}$ ends up being positive, highlighting the importance of relying on experimental variation for accurate estimates.

\begin{table}[!htb] \centering 
\caption{Using experimental variation to quantify the effect of pod length and frequency.}
  \label{table:pod_len_freq} 
\begin{tabular}{lcc}
   \tabularnewline \midrule \midrule
   Dependent Variables:    & Total Hours    & Active Days\\  
   Model:                  & (1)            & (2)\\  
   \midrule
   Pods of Len 1 Per hour ($\beta_1$)     & -1.966$^{***}$ & -1.654$^{***}$\\   
                           & (0.2241)       & (0.1262)\\   
   Pods of Len 2 Per hour ($\beta_2$)     & -4.116$^{***}$ & -3.715$^{***}$\\   
                           & (0.2443)       & (0.1397)\\   
   Constant                & 107.3$^{***}$  & 106.2$^{***}$\\   
                           & (0.6652)       & (0.3757)\\   
   \midrule
   Test of difference ($2\cdot \beta_1 - \beta_2 = 0)$      & 0.1845         & 0.4073$^{**}$\\  
                         & (0.3055)       & (0.1704)\\  
   \midrule
   Observations            & 34,390,962     & 34,390,962\\  
   F-test (1st stage), Pods of Len 1 Per hour & 97,510.4       & 97,510.4\\  
   F-test (1st stage), Pods of Len 2 Per hour & 326,039.2      & 326,039.2\\ 
   \midrule \midrule
   \multicolumn{3}{l}{\emph{Clustered (listener level) standard-errors in parentheses}}\\
   \multicolumn{3}{l}{\emph{Signif. Codes: ***: 0.01, **: 0.05, *: 0.1}}\\
\end{tabular}
\end{table} 

\subsection{Impact on Demand for the Ad-Free Subscriptions} \label{sect:subscriptions}

For platforms with a freemium model, understanding the relationship between their free, ad-supported version and their paid, ad-free version is crucial for optimizing revenue streams. The free version can serve either as a substitute or a complement to the paid version, depending on how users perceive and interact with the service. When the free version acts as a complement, it allows users to explore the platform's features and build a habit, potentially leading to a paid subscription later on, similar to a trial period. Conversely, when the free version serves as a substitute, increasing the ad load may nudge users to upgrade to the paid version to avoid the disruption caused by advertisements.

The impact of ad load on subscriptions is an important consideration for Pandora's business model, as it directly affects the company's revenue streams. Pandora offers both a free, ad-supported version and a paid, ad-free version called Pandora One, which was priced at \$4.99 per month during the experiment period. The effect of an increase in ad load on subscriptions to Pandora One is a priori unclear. Increasing ad load might reduce user engagement with the app, hindering the development of a habit stock that fosters loyalty and willingness to pay for a subscription. However, it might also make the free version less attractive and lead users to upgrade to Pandora One.

\begin{table}[!htb] \centering 
\caption{IV, effect of ad load per hour on listener churn and Pandora One subscription.}
  \label{table:iv-churn-p1} 
\begin{tabular}{lcc}
   \tabularnewline \midrule \midrule
   Dependent Variables: & Subscriber At End & Listener Churn\\  
   Model:               & (1)               & (2)\\  
   \midrule
   Ads Per Hour         & 0.00145$^{***}$   & 0.00336$^{***}$\\   
                        & (0.00004)         & (0.00018)\\   
   Subscriber At Start  & 0.34949$^{***}$   & -0.00632$^{***}$\\   
                        & (0.00081)         & (0.00085)\\   
   Constant             & 0.00464$^{***}$   & 0.51752$^{***}$\\   
                        & (0.00015)         & (0.00069)\\   
   \midrule
   Observations         & 34,390,962        & 34,390,962\\  
   F-test (1st stage) & 157,980.3         & 157,980.3\\  
   \midrule \midrule
   \multicolumn{3}{l}{\emph{Clustered (listener level) standard-errors in parentheses}}\\
   \multicolumn{3}{l}{\emph{Signif. Codes: ***: 0.01, **: 0.05, *: 0.1}}\\
\end{tabular} 
\end{table}

We measure this effect in Table \ref{table:iv-churn-p1}, where the binary outcome is the listener's subscription status on the last day of the experiment. This 2SLS regression measures how this outcome depends on ads per hour, with the additional covariate of subscription status at the beginning of the experiment. Recall that we study only the set of individuals who engaged in at least some ad-supported listening during the experiment, so any listeners in this dataset who were subscribers at the beginning of the experiment must have allowed their subscriptions to lapse at some point during the experiment. Even for this set of individuals, we find that those who had paid subscriptions on the first day of the experiment were 35 percentage points more likely to be subscribers at the end than those who didn't subscribe at the beginning of the experiment.

We observe that for each one ad per hour increase, there is a 0.34\% decrease in the probability of listening at all in the final month, and a 0.14 percentage-point increase in the probability of being a paid subscriber at the end of the experiment. Given the subscription fee of \$4.99 per month, this translates to an increase in monthly subscription revenue of approximately 0.75 cents per listener for each additional ad per hour (minus payment-processing costs). These findings provide insights into how ad load affects listener retention, subscription rates, and potential revenue changes. However, our research agreement with Pandora does not allow us to conduct or discuss a comprehensive cost-benefit analysis that would require additional proprietary data on advertising revenues and other business metrics.

This comparison highlights the potential trade-off between increasing ad load, driving subscription rates, and retaining listeners. Although increasing ad load may encourage some users to upgrade to Pandora One, it also drives away a larger proportion of listeners who choose to stop using the service entirely. While it might appear that using ad load could be an effective way to nudge users to upgrade, the increase in churn indicates that uniformly increasing ad load might be too costly for the platform.

In Online Appendix A, we demonstrate that using observational data can be misleading for this purpose as well. While observational data shows that increasing ad load increases churn, it also suggests that increasing ad load lowers subscription rates, implying that ad-free and paid versions are complements. This result can be rationalized with economic theory: higher ad loads might make the experience with the ad-supported version worse, potentially preventing users from building habit stock or leading them to make negative inferences about the platform as a whole, thus reducing the likelihood of subscription. Although this reasoning seems plausible, our experimental results show that this inference is incorrect. This discrepancy highlights the importance of having experimental variation for accurate estimates. We refer readers to Online Appendix A for a detailed comparison of observational and experimental results.

This finding emphasizes the importance of carefully balancing the ad load to optimize overall revenue, considering both subscription and advertising revenues while minimizing listener attrition. Our analysis also reveals significant heterogeneity in how different age groups respond to increased ad load (detailed in Online Appendix B), which suggests potential value in personalizing ad load strategies. In a complementary study, \citet{goli2021personalized} use a later Pandora experiment to leverage a large-scale machine learning approach with extensive pre-treatment feature collection to develop and evaluate methods for personalizing ad load based on user characteristics, demonstrating how targeted strategies could enhance subscription profits without decreasing overall advertising profits. Their work shows how data-driven personalization techniques could help mitigate the costly effects of uniform increases in ad load that we document here.

\section{Conclusion} \label{sect:conclusion}

We reported on the results of an extensive experiment undertaken by Pandora to trace out its consumers' demand curve for Pandora listening as a function of the number of audio ads served. With more than 30 million listeners experiencing one of nine different randomized treatments consistently for a period of 21 months, we have obtained a robust estimate of a demand curve. We found that the number of hours listened is a monotonically decreasing function of the price (number of ads per hour). We find that, conditional on the same overall ad load, consumers show a preference for more frequent but shorter ad breaks (pods). Our results suggest that the majority of the reduction in listening hours is likely due to a decrease in the number of active days per listener and the likelihood of being an active listener at all, rather than a reduction in the number of hours per active day. This indicates that the primary effect of increased ad load is on users' decisions to initiate listening sessions, with a smaller impact on the duration of those sessions once started.

Importantly, we found that the long-run elasticity of demand with respect to ad load is three times larger than the short-run elasticity. This finding underscores the crucial need for extended observation periods in understanding ad load effects. Had we run an experiment for just a month or two, we could have underestimated the true long-run effects by a factor of 3.

We demonstrate the importance of using an experiment rather than relying on endogenous observational data. We explored various observational methods, including OLS, OLS with fixed effects and controls, instrumental variables, and panel regressions. Each of these methods led to biased estimates of the causal effects of ad load on listener behavior. For instance, simple OLS and OLS with fixed effects significantly overestimated the effect of ad load on total hours listened. The panel regression, while closer to the experimental estimates for active days, still overestimated the effect on total hours by a factor of 3. These discrepancies highlight the value of experimental data in accurately measuring consumer responses to changes in ad load.

We also explored heterogeneity in the impact of increased ad load on subscriptions and churn across age groups, as detailed in Online Appendix B. Older listeners (55+) were twice as likely as younger listeners (13-24) to respond to higher ad load by subscribing to the ad-free service, while younger listeners were more prone to leave the platform entirely. These findings highlight the potential for targeted ad load strategies to improve user retention and subscription revenue. Using a subsequent experiment with a richer set of pre-treatment features, \citet{goli2021personalized} employ machine learning methods to estimate conditional average treatment effects (CATEs), demonstrating how personalized ad load policies can optimize subscription and advertising revenues.

We acknowledge a potential limitation in our study design. By measuring total listening across all devices while only manipulating ad load on mobile devices (which constitute about 80\% of listening), our elasticity estimates may be somewhat attenuated if users switch from mobile devices to non-mobile devices like desktop browsers. However, the consistency of our results across a wide range of ad loads suggests that this attenuation, if present, is likely small and uniform across treatments. The shape of the observed demand curve (see Figure~\ref{fig:hours-overall}) further supports this view, as significant substitution to non-mobile devices would likely result in a flatter demand curve, particularly at higher ad loads. While we cannot precisely quantify this effect due to limitations in device-specific tracking at the time of the experiment, our long-term approach provides some reassurance. For substitution to non-mobile devices to significantly impact our results, it would need to be both substantial and persistent over the 21-month experiment period, which seems unlikely given the observed relationship between ad load and listening behavior (Figure~\ref{fig:hours-overall}). 
\footnote{Another potential source of bias in our experiment is word-of-mouth effects, which could cause spillovers between treatment and control groups, potentially leading to underestimation of ad load sensitivity. However, since ad load is an implicit cost experienced over time rather than an explicit posted price, users are less likely to directly compare experiences, mitigating this issue to some extent. While we believe such spillover effects to be negligible for existing Pandora listeners in our study, we acknowledge this possibility.}

Our estimates are quite precise, with the slope coefficients in our main outcome regressions having $t$-statistics of approximately 20 (i.e., point estimates five times larger than the width of their 95\% confidence intervals). We also saw that the effects of a change in ad load take at least a year to be fully realized, demonstrating the importance of our having run a long-term experiment. The long-run elasticity of demand with respect to ad load is three times larger than the short-run elasticity, highlighting the crucial need for extended observation periods in understanding ad load effects.

These results go a long way towards helping us understand the science of two-sided markets, as we have managed to describe the behavior of one side of Pandora's advertising business in great detail. Our findings also underscore the importance of experimental data in accurately understanding platform dynamics. We demonstrated that various observational methods, including panel data techniques and the inclusion of granular fixed effects, often lead to biased estimates of the causal effects of ad load on listener behavior, sometimes even yielding incorrect directional effects.

The results are also quite relevant to Pandora's business, as the firm has gained a much better understanding of its listeners through this experiment. Prior to the experiment, Pandora had almost no data on how sensitive its ad-supported listeners were to the level of advertising. Armed with the demand-curve estimates presented here, the company has much better information for deciding appropriate levels of advertising for its audience. Moreover, our findings on ad scheduling preferences provide valuable insights for improving user experience without necessarily reducing overall ad load.  The evidence that listeners prefer more frequent but shorter ad breaks presents an opportunity to enhance user experience while maintaining revenue. The results on heterogeneous treatment effects by age are also quite valuable for understanding listeners' decisions to pay for ad-free subscriptions and could potentially be leveraged to personalize ad load and optimize user experience and revenues.

This study not only informs Pandora's business strategy but also contributes to experimental research in digital economics. While experimentation has become increasingly common in the digital economy, studies of this scale and duration remain relatively rare. It demonstrates the value of long-term, large-scale experimentation in uncovering causal relationships and guiding business decisions in the digital age. Our work highlights the importance of extended observation periods in capturing the full effects of interventions like changes in ad load. We hope this work encourages more firms to consider conducting similarly extensive experiments, as they can provide insights into consumer behavior and platform dynamics that short-term studies or observational data alone might miss. Such rigorous, long-term experimentation can be crucial for optimizing strategies in the complex ecosystems of digital platforms and advancing our understanding of two-sided markets.

{
\singlespacing
\newpage
\bibliographystyle{apalike2}
\bibliography{main}
}

\pagebreak 

\appendix
\section*{Online Appendix A - Business Implications Using Observation Data }
\renewcommand{\thetable}{A\arabic{table}}
\setcounter{table}{0}

In Section~\ref{sec:business}, we discussed insights derived from experimental variation regarding ad scheduling and the substitution/complementarity relationship between the free and paid versions of the service. In this section, we demonstrate that relying on observational variation to answer these questions would have led to misleading conclusions.

We begin by re-estimating the counterpart to model~\eqref{eq:simp_iv_app} using data from the control condition. Table~\ref{table:ols_pod_len_freq} presents the estimates for OLS and OLS with fixed effects and controls, similar to the analyses in Section~\ref{sec:OLS_FE}. The results from the simple OLS without fixed effects and controls, shown in columns (1) and (3), yield a positive coefficient for $\beta_2$ for both Total Hours and Active Days. If interpreted causally, this would be misleading, as it implies that pods of length 2 increase user consumption and listeners derive consumption utility from pods of length 2 but not from those of length 1.

\begin{table}[ht]
\centering
\caption{Using observational variation in the control condition to quantify the effect of pod length and frequency.}
\label{table:ols_pod_len_freq}
\resizebox{0.8\textwidth}{!}{%
\begin{tabular}{lcccc}
   \tabularnewline \midrule \midrule
   Dependent Variables: & \multicolumn{2}{c}{Total Hours} & \multicolumn{2}{c}{Active Days}\\
   Model:                  & (1)            & (2)            & (3)            & (4)\\  
   \midrule
   \# Pods of Len 1 Per Hour ($\beta_1$)     & -6.821$^{***}$ & -9.298$^{***}$ & -3.797$^{***}$ & -6.546$^{***}$\\   
                           & (0.0450)       & (0.2656)       & (0.0136)       & (0.0128)\\   
   \# Pods of Len 2 Per Hour ($\beta_2$)     & 1.393$^{***}$  & -4.018$^{***}$ & 4.490$^{***}$  & -1.937$^{***}$\\   
                           & (0.0883)       & (0.7837)       & (0.0355)       & (0.0309)\\   
   Constant                & 115.4$^{***}$  &                & 106.2$^{***}$  &   \\   
                           & (0.2202)       &                & (0.0654)       &   \\   
   \midrule
   Pre-treatment controls &  & X &  & X \\
   Zip code FE                 &                & X            &                & X\\  
   Listener age FE              &                & X            &                & X\\  
   \midrule
   Observations            & 18,097,258     & 18,097,258     & 18,097,258     & 18,097,258\\  
   \midrule \midrule
   \multicolumn{5}{l}{\emph{Clustered (listener level) standard-errors in parentheses}}\\
   \multicolumn{5}{l}{\emph{Signif. Codes: ***: 0.01, **: 0.05, *: 0.1}}\\
\end{tabular}%
}
\end{table}

This positive coefficient for $\beta_2$ might be due to the fact that users who are in higher demand by advertisers are more likely to be served more ad pods of length 2, as it is easier for the system to find relevant ads for them. Simultaneously, these same demographics might also listen to Pandora for longer hours, leading to spurious correlation and a positive coefficient for $\beta_2$. When we control for zip code and listener age fixed effects, along with pre-treatment consumption features (columns (2) and (4) of Table~\ref{table:ols_pod_len_freq}), the coefficient for $\beta_2$ becomes much more similar to the one measured in Table~\ref{table:pod_len_freq}, which uses experimental variation. However, the coefficient for $\beta_1$ is now severely biased, and indeed $|\beta_1| > |\beta_2|$, which would imply that users prefer a pod of length 2 to a pod of length 1, contradicting the findings from the experimental variation.

These results highlight the challenges of using observational data to infer the optimal ad scheduling mechanism, given the complex confounding factors that are present. The inconsistencies and biases in the estimates obtained from observational data underscore the importance of relying on experimental variation to accurately measure the causal effects of ad scheduling on user behavior and preferences. 

We then repeat this analysis to measure the effect of ad load on churn and subscriptions using the observational data in the control condition. The estimates for the effect of ad load on subscription status and listener churn are presented in columns (1)-(2) and (3)-(4) of Table~\ref{table:OLS_churn_sub}. Similar to our analysis above, we add zip code and listener age fixed effects in columns (2) and (4). While the coefficient for the effect on churn has the correct sign (i.e., the same sign as in Table~\ref{table:iv-churn-p1}, which uses experimental variation) and bias reduces as we add the fixed effect, the coefficient for effect on subscription is negative (has the wrong sign) and the results become even more biased as we add finer level fixed effects. At first glance, the results seem plausible: churn increases because of higher ad load, and it could be that higher ad load reduced engagement with the ad-supported service, making it less likely for users to upgrade to the paid service because of less habit stock or poor experience with the ad-supported version and the platform as a whole. In other words, the ad-supported and paid services were complements. However, as we know from the experimental variation and the results in Table~\ref{table:iv-churn-p1}, this is not correct, and higher ad load actually increases subscriptions to the paid service. These results once again demonstrate that using observational data can be misleading and lead to suboptimal business decisions.

\begin{table}[ht]
\centering
\caption{Using observational variation in the control condition to quantify the effect of ad load per hour on listener churn and Pandora One subscription.}
\label{table:OLS_churn_sub}
\resizebox{0.8\textwidth}{!}{%
\begin{tabular}{lcccc}
   \tabularnewline \midrule \midrule
   Dependent Variables: & \multicolumn{2}{c}{Subscriber At End} & \multicolumn{2}{c}{Listener Churn}\\
   Model:                 & (1)              & (2)              & (3)             & (4)\\  
   \midrule
   Ads Per Hour           & -0.00097$^{***}$ & -0.00145$^{***}$ & 0.01410$^{***}$ & 0.02387$^{***}$\\   
                          & (0.00001)        & (0.00001)        & (0.00005)       & (0.00005)\\   
   Subscriber At Start    & 0.34815$^{***}$  & 0.32678$^{***}$  & 0.00004         & 0.19038$^{***}$\\   
                          & (0.00111)        & (0.00119)        & (0.00117)       & (0.00214)\\   
   Constant               & 0.01340$^{***}$  &                  & 0.47850$^{***}$ &   \\   
                          & (0.00004)        &                  & (0.00021)       &   \\   
   \midrule
   Pre-treatment controls &                  & X                &                 & X\\  
   Zip code FE                &                  & X              &                 & X\\  
   Listener age FE              &                  & X              &                 & X\\  
   \midrule
   Observations           & 18,097,258       & 18,097,258       & 18,097,258      & 18,097,258\\  
   \midrule \midrule
   \multicolumn{5}{l}{\emph{Clustered (listener level) standard-errors in parentheses}}\\
   \multicolumn{5}{l}{\emph{Signif. Codes: ***: 0.01, **: 0.05, *: 0.1}}\\
\end{tabular}%
}
\end{table}

\clearpage

\section*{Online Appendix B - Heterogeneous Effects of Ad Load on Subscriptions and Churn Across Age Groups}
\renewcommand{\thetable}{B\arabic{table}}
\setcounter{table}{0}

In this appendix, we examine whether the effects of ad load on subscription and churn vary systematically with user age. Age is a readily available user characteristic that likely correlates with factors relevant to these decisions, such as disposable income and familiarity with alternative platforms. Understanding such variation could reveal whether there is potential value in more nuanced ad load strategies. We analyze four age groups (13-17, 18-24, 25-54, 55+), estimating separate treatment effects while controlling for initial subscription status.\footnote{We initially allowed the coefficients to vary both by age and gender, but found that gender differences were negligible. While developing personalization strategies is beyond the scope of this study, our goal is to understand if meaningful heterogeneity exists in treatment effects. For a comprehensive approach to ad load personalization, see \citet{goli2021personalized}, who use machine learning methods to detect heterogeneous treatment effects using a later large-scale field experiment at Pandora and optimize personalized ad load strategies, leveraging a rich set of pre-treatment features including zip-level income, pre-treatment consumption patterns, and other user characteristics.}

\begin{model}[Effect of Ad Load Across Age Group]
Let $y_i$ denote total listening hours or days in the final month. Let \textbf{$Demo_i$} denote a vector of dummy variables corresponding to the demographic groups a given individual belongs to. The Demographic-specific regression model we run is given by:
\begin{align}\label{eq:demo_iv}
y_i & = \textbf{Demo}_i^T \boldsymbol{\beta}_0 + \widehat{\frac{ads}{hours}}_i \cdot \textbf{Demo}_i^T \boldsymbol{\beta}_1 + \varepsilon_i 
\end{align}
\end{model}

\begin{table}[!htbp] \centering 
\caption{IV effect of ad load per hour on listener churn and Pandora One subscription by demographics.}
  \label{table:iv-churn-p1-by-age} 
\begin{tabular}{lcc}
   \tabularnewline \midrule \midrule
   Dependent Variables:                          & Subscriber At End & Listener Churn\\  
   Model:                                        & (1)               & (2)\\  
   \midrule
   Ads Per Hour x Age 13-17     & 0.00059$^{***}$   & 0.00143\\   
                                                 & (0.00017)         & (0.00178)\\   
   Ads Per Hour x Age 18-24     & 0.00088$^{***}$   & 0.00325$^{***}$\\   
                                                 & (0.00006)         & (0.00036)\\   
   Ads Per Hour x Age 25-54     & 0.00165$^{***}$   & 0.00361$^{***}$\\   
                                                 & (0.00005)         & (0.00020)\\   
   Ads Per Hour x Age 55+       & 0.00212$^{***}$   & 0.00220$^{*}$\\   
                                                 & (0.00031)         & (0.00116)\\   
   Age 13-17                      & 0.00077$^{***}$   & 0.60886$^{***}$\\   
                                                 & (0.00029)         & (0.00313)\\   
   Age 18-24                      & 0.00189$^{***}$   & 0.54079$^{***}$\\   
                                                 & (0.00024)         & (0.00142)\\   
   Age 25-54                      & 0.00581$^{***}$   & 0.48433$^{***}$\\   
                                                 & (0.00022)         & (0.00086)\\   
   Age 55+                        & 0.00950$^{***}$   & 0.54610$^{***}$\\   
                                                 & (0.00083)         & (0.00308)\\   
   Subscriber At Start                           & 0.34799$^{***}$   & 0.01072$^{***}$\\   
                                                 & (0.00081)         & (0.00087)\\   
   \midrule
   Observations                                  & 34,390,962        & 34,390,962\\  
   F-test (1st stage), Ads Per Hour x Age 13-17 & 5,318.9           & 5,318.9\\  
   F-test (1st stage), Ads Per Hour x Age 18-24 & 40,592.7          & 40,592.7\\  
   F-test (1st stage), Ads Per Hour x Age 25-54 & 62,260.4          & 62,260.4\\  
   F-test (1st stage), Ads Per Hour x Age 55+   & 9,358.6           & 9,358.6\\   
   \midrule \midrule
   \multicolumn{3}{l}{\emph{Clustered (listener level) standard-errors in parentheses}}\\
   \multicolumn{3}{l}{\emph{Signif. Codes: ***: 0.01, **: 0.05, *: 0.1}}\\
\end{tabular}
\end{table}

The estimates from model~\eqref{eq:demo_iv} are presented in Table \ref{table:iv-churn-p1-by-age}. The first column shows that both the baseline probability and the marginal change in the probability of holding an ad-free subscription at the end of the experiment are monotonically increasing in age. As before, we include subscription status at the beginning of the experiment as a covariate. We see that listeners over 55 years old are twice as likely as listeners between the ages of 13 and 24 (marginal impact of 0.21\% versus 0.09\% or less) to react to an increase in ad load by paying for the ad-free service. To the extent that older people have more disposable income, these results are consistent with a positive income elasticity of demand for the ad-free subscription.

For comparison, in the second column of Table \ref{table:iv-churn-p1-by-age}, we provide a related set of heterogeneous treatment effects, this time for the probability of leaving Pandora altogether. Pandora generally considers a listener to have ``churned'' away from the service if that person has not listened to the service for two months. We therefore define a binary outcome variable equal to 1 if the listener did not listen at all during the last two months of the experiment. The treatment effects display less heterogeneity than in the subscription results. Comparing the two columns of Table \ref{table:iv-churn-p1-by-age} shows us that listeners aged 18-24 are much more likely to react to increased ad load by leaving Pandora than by paying for an ad-free subscription (0.33\% versus 0.09\% for each additional ad per hour). By contrast, listeners over 55 years old are nearly as likely to react by switching to an ad-free subscription as by leaving Pandora altogether (0.21\% versus 0.22\% for each additional ad per hour).

While our analysis reveals significant heterogeneity across age groups, it is important to note that we do not believe age itself is causing these differences in behavior. Rather, age serves as a proxy for various factors that may influence elasticities, such as disposable income, familiarity with technology, and engagement with the platform. Digital content platforms like Pandora can be considered experience goods, where users' responses to changes such as increased ad load may vary depending on their level of engagement with the platform. More committed users might be more willing to upgrade to an ad-free subscription when faced with increased ad load, possibly due to a more personalized listening experience from a recommender system well-tuned to their taste.

Our current analysis does not explicitly account for these potential sources of heterogeneity beyond age groups, which presents an opportunity for future research. However, the observed age-based heterogeneity likely captures some of these differences. For instance, the stark difference in churn rates versus subscription rates between younger and older listeners (e.g., 18-24 age group showing a 2.7:1 ratio of churn to subscription, compared to nearly 1:1 for the 55+ group) suggests that younger users may have lower switching costs, possibly due to greater familiarity with alternative platforms, while older users may have higher disposable income to afford subscriptions.

These results are interesting because they suggest that for every user converted to the paid version in the 55+ age group, only one user left the platform. However, this ratio is almost 3 to 1 for users in the 18-24 age group. These differences likely reflect a combination of (a) the higher disposable income of the older age group and (b) the availability of or familiarity with alternative platforms among younger listeners, rather than age itself being the causal factor.

\clearpage

\section*{Online Appendix C - Adjusted Standard Errors for Normalized Outcomes}
\renewcommand{\thetable}{C\arabic{table}}
\setcounter{table}{0}

In Section \ref{sec:measurement}, we normalized our outcome variables by dividing by the control group mean to protect confidentiality while maintaining interpretability. While our control group contains millions of users, making the standard errors around these control means quite tight, this normalization introduces an additional source of variation that should be accounted for in our standard error calculations. As our setup exactly parallels that in~\citet{goli2021personalized}, we adopt their analysis from Online Appendix G here, reproducing their derivations here for completeness with notation adapted to our context. The calculations that follow are mathematically identical to theirs, as we face the same statistical problem of conducting inference on estimates normalized by a control group mean.

Formally, for our analyses in Section \ref{sec:measurement}, we normalized outcomes using:

\begin{equation}
\tilde{y}_i = 100 \cdot \frac{y_i}{\frac{\sum_{j \in \mathcal{C}} y_j}{\mathcal{N}_\mathcal{C}}},
\label{eq:scale_appendix}
\end{equation}
where $y_i$ is an outcome of interest (hours, active days, or active user), and $\tilde{y}_i$ is its normalized counterpart. $\mathcal{C}$ represents the set of users in the control condition and $\mathcal{N}_\mathcal{C}$ is the number of users in control. Our instrumental variables regression on these normalized outcomes takes the form:
\begin{equation}
\tilde{y}_i = \alpha + \beta\cdot\mathcal{A}_i + \epsilon_i,  
\label{eq:iv_reg_appendix}
\end{equation}
Without normalization, the regression would be:
\begin{equation}
{y}_i = \alpha + \beta^*\cdot\mathcal{A}_i + \epsilon^*_i,  
\label{eq:iv_reg_true}
\end{equation}
The reported $\beta$ in \eqref{eq:iv_reg_appendix} is therefore:

\[ \beta = \frac{\beta^*}{m_y},\]
where $m_y = \frac{\sum_{j \in \mathcal{C}} y_j}{\mathcal{N}_\mathcal{C}}$ represents the average outcome in the control group. To conduct inference on $\beta$, we can estimate $\beta^* \sim \mathcal{N}(\mu_{\beta^*}, \sigma_{\beta^*})$ and $m_{y} \sim \mathcal{N}(\mu_m , \sigma_m)$. Following the delta method for $g(x,y) = \frac{x}{y}$:

\[\frac{\partial}{\partial \beta} g(\beta, m) = \frac{1}{m},\]

\[\frac{\partial}{\partial  m} g(\beta, m) = -\frac{\beta}{m^2}.\]

Using the Taylor expansion:
\[ \mathbf{E} \left[ g(\beta^*,m_y)\right] \approx \frac{\mu_{\beta^*}}{\mu_m},\]
and
\begin{equation}
\begin{aligned}
    Var(g(\beta^*,m_y)) &= Var\left(\frac{\beta^*}{m_y}\right)\overset{(1)}{\approx} \left(\frac{\mu_{\beta^*}}{\mu_m}\right)^2 \left( \frac{\sigma^2_{\beta^*}}{\mu^2_{\beta^*}} + \overset{\text{adjustments}}{\overbrace{\frac{\sigma^2_{m}}{\mu^2_{m}} - 2 \frac{Cov(\beta^*,m_y)}{\mu_m \cdot \mu_{\beta^*}}}} \right)\\
    &\overset{(2)}{\leq} \left(\frac{\mu_{\beta^*}}{\mu_m}\right)^2 \left( \frac{\sigma^2_{\beta^*}}{\mu^2_{\beta^*}} +  \frac{\sigma^2_{m}}{\mu^2_{m}} +  2 \left| \frac{\sigma_{\beta*} \cdot \sigma_{m} }{\mu_m \cdot \mu_{\beta^*}}\right|   \right),
\label{eq:se_adjustment}
\end{aligned}
\end{equation}
where (1) follows from the Taylor expansion of $g(\cdot,\cdot)$ around $(\beta^*, m_y)$ and the delta method, and (2) is a consequence of the triangle inequality and the Cauchy-Schwarz inequality. To avoid dealing with the covariance of $\beta^*$ and $m_y$, we bound it above using the Cauchy-Schwartz inequality.

This calculation provides a conservative estimate for the standard error of $\beta$. If we were to treat the normalizing factor as a constant rather than an estimate, we would omit the terms labeled as ``adjustments'' in equation \eqref{eq:se_adjustment}. Given our large control group of 18 million listeners, $\sigma_m$ is quite small. The adjustment accounts for approximately 2 to 10\% of the unadjusted standard error across our various outcomes, which translates to less than 1\% of the estimate's magnitude. Table~\ref{table:simple-iv-wquad-adjusted} presents the counterpart to Table~\ref{table:simple-iv-wquad}, with standard errors adjusted to account for the uncertainty in estimating the control means using the correction outlined above. Specifically, the standard errors for Total Hours and Active Days increase from 0.1157 to 0.1267 and from 0.0662 to 0.0673, respectively. These adjustments represent less than 1\% of their corresponding point estimates of -2.082 and -1.911.

\begin{table}[!htbp] 
\centering 
\caption{The counterpart of treatment effect estimates in Table \ref{table:simple-iv-wquad} after adjusting for uncertainty in the control group means (normalization factor) in standard errors.}
\label{table:simple-iv-wquad-adjusted} 
\begin{tabular}{@{\extracolsep{5pt}}lcc} 
   \tabularnewline \midrule \midrule
   Dependent Variables: & Total Hours    & Active Days\\  
                  & (1)            & (2)\\  
   \midrule
   Ad Per Hour     & -2.081$^{***}$ & -1.911$^{***}$\\   
                        & (0.1267)       & (0.0673)\\   
   \midrule
   Observations         & 34,390,962     & 34,390,962\\  
   F-test (1st stage) & 157,839.4      & 157,839.4\\  
   \midrule \midrule
   \multicolumn{3}{l}{\emph{Clustered (listener level) standard-errors in parentheses}}\\
   \multicolumn{3}{l}{\emph{Signif. Codes: ***: 0.01, **: 0.05, *: 0.1}}\\
\end{tabular} 
\end{table}

\clearpage

\section*{Online Appendix D - Decomposing the Treatment Effect using Panel Structure and Experimental Data}
\renewcommand{\thetable}{D\arabic{table}}
\setcounter{table}{0}

In Section~\ref{sec:decompose}, our decomposition analysis suggested that the majority of the decline in listening hours comes from decreased engagement on the extensive margin rather than shorter listening sessions. However, this analysis faces two potential selection issues. First, increased ad load leads to churn off the platform, which means the set of users who remain active in any given period may systematically differ in their listening patterns. Second, more ad-sensitive users may be more likely to churn early, leaving a selected sample of less sensitive users and potentially leading to underestimation of the effect on consumption conditional on activity.

To provide additional evidence supporting our findings and further explore these effects while accounting for pre-treatment differences across listeners, we estimate a panel specification that combines both experimental variation and pre-post comparisons. We restrict our analysis to listeners who were active in May 2014 (pre-treatment) across all experimental conditions who had exposure to treatment in the first six months of the experiment. We create a balanced panel with two observations per listener -- one before treatment and one in the final month. We then examine four key margins of adjustment: total hours, activity (probability of listening at all), active days conditional on activity, and hours per active day during the final month. While users may still differentially exit the sample in response to treatment, this specification controls for cross-sectional differences in listening behavior prior to the experiment, which likely addresses a substantial portion of the selection concerns. By leveraging within-listener experimental variation while controlling for pre-existing differences across users, it provides complementary evidence supporting our main findings. The model can be specified as follows:

\begin{model}[Panel IV with Experimental Variation]
Let $y_{it}$ denote the outcome of interest for listener $i$ in period $t \in \{pre,post\}$, where $pre$ refers to May 2014 (before the experiment) and $post$ refers to the final month. The regression model we run is given by:
\begin{align}\label{eq:panel_iv}
y_{it} & = \delta_i + \tau_t + \widehat{\left(\frac{ads}{hours}\right)}_{it} \cdot \beta_1 + \varepsilon_{it} \\
\left(\frac{ads}{hours}\right)_{it} & = \delta_i + \tau_t + (TG_i \times Post_t)' \cdot \gamma + \eta_{it} \nonumber
\end{align}
where $\delta_i$ and $\tau_t$ are listener and period fixed effects respectively, $TG_i$ is a vector of treatment group indicators, and $Post_t$ is an indicator for the post-treatment period. The interaction terms $TG_i \times Post_t$ serve as instruments for the realized ad load.
\end{model}

\begin{table}[!htbp]
\centering
\caption{Effect of audio ads per hour on listening outcomes, estimated using a panel data specification. The analysis focuses on four margins: total hours, activity, active days (conditional on activity), and hours per active day during the final month. This specification uses two observations per listener and is restricted to users who were active in May 2014 (pre-treatment) across all experimental conditions.}
\label{table:breakdown_IV_panel}
\begin{tabular}{@{\extracolsep{5pt}}l*{6}{p{2cm}}}
\\[-1.8ex]\hline
\hline \\[-1.8ex]
Dependent Variables: & Total Hours & Active & Days/Active Listener & Hours/Active Day\\
& (1) & (2) & (3) & (4)\\
\midrule
Ad Per Hour & -2.187$^{***}$ & -1.065$^{***}$ & -0.9664$^{***}$ & -0.4637$^{***}$\\
& (0.1765) & (0.0533) & (0.0635) & (0.0875)\\
\midrule
Observations & 25,128,520 & 25,128,520 & 13,724,840 & 13,724,840\\
F-test (1st stage) & 161,430.1 & 161,430.1 & 181,303.1 & 181,303.1\\
\midrule \midrule
\multicolumn{5}{l}{\emph{Clustered (listener level) standard-errors in parentheses}}\\
\multicolumn{5}{l}{\emph{All regressions include listener and period fixed effects}}\\
\multicolumn{5}{l}{\emph{Signif. Codes: ***: 0.01, **: 0.05, *: 0.1}}\\
\end{tabular}
\end{table}

The results from the panel specification (Table \ref{table:breakdown_IV_panel}) are remarkably similar to those from the simple IV specification (Table \ref{table:iv-hours-tab}), despite using a different sample and methodology. The panel estimates show that one additional ad per hour reduces mean listening time by 2.187\% (SE: 0.176\%), compared to 2.082\% (SE: 0.116\%) in the simple IV specification. Similarly, the effect on active days conditional on activity (-0.966\% vs -0.955\%), and hours per active day (-0.464\% vs -0.403\%) are nearly identical across both specifications.

However, this analysis does not fully address potential selection driven by differential churn of more ad-sensitive users over time. If more ad-sensitive users exit the sample earlier, our final month estimates of consumption effects conditional on activity could be attenuated. To investigate this possibility, we re-estimate our panel specification comparing pre-treatment outcomes to those in January 2015, six months into the experiment. At this point, the treatment effects had reached roughly half their eventual steady-state magnitude (see Figure \ref{fig:temporal_estimates}), providing an intermediate snapshot of user responses before full adjustment had occurred.

If our final month results were primarily driven by selection of less sensitive users remaining in the sample, we would expect to see a larger effect on hours per active day relative to total hours in January 2015, before more sensitive users had churned. The results in Table \ref{table:breakdown_IV_panel_jan_2015} do not support this hypothesis. While the total effect on listening (-0.974\%) is about half the final steady-state effect (-2.187\%), the ratio of the effect on hours per active day to total hours is even smaller than in our final month analysis (0.045/0.974 vs 0.464/2.187). Instead, the intermediate effect is driven primarily by reductions in the extensive margin -- both the probability of being active and the number of active days conditional on activity.

While selection issues cannot be fully ruled out in this setting, these analyses using panel structure and experimental variation provide additional insight into the decomposition of effects presented in Section~\ref{sec:decompose}. The relatively small effects on hours per active day at both intermediate and final time periods, combined with our main decomposition results in Section~\ref{sec:decompose}, collectively suggest that the reduction in listening hours comes primarily through decreased frequency of listening rather than shortened session lengths.

Additionally, these analyses provide useful evidence regarding the robustness of our findings to the continuous enrollment design of the experiment.\footnote{As discussed in Section~\ref{sect:design}, new listeners joining Pandora were assigned to treatment using the same randomization throughout the experiment period.} The panel specification uses a fixed set of listeners who were active in May 2014, while our main analysis includes all users, including those who joined during the experiment period. The consistency of effects across these different samples suggests that the timing of enrollment and differences in treatment exposure duration are not major drivers of our key findings regarding the gradual adjustment of listening behavior and the long-run versus short-run elasticities documented in Section~\ref{sec:measurement}.

\begin{table}[!htbp]
\centering
\caption{Effect of audio ads per hour on listening outcomes, estimated using a panel data specification. The analysis focuses on four margins: total hours, activity, active days (conditional on activity), and hours per active day during January 2015 (6 months into the experiment). This specification uses two observations per listener and is restricted to users who were active in May 2014 (pre-treatment) across all experimental conditions.}
\label{table:breakdown_IV_panel_jan_2015}
\begin{tabular}{@{\extracolsep{5pt}}l*{6}{p{2cm}}}
\\[-1.8ex]\hline
\hline \\[-1.8ex]
   Dependent Variables: & Total Hours     & Active          & Days/Active Listener & Hours/Active Day\\  
   Model:               & (1)             & (2)             & (3)                  & (4)\\  
   \midrule
   Ad per Hour            & -0.9741$^{***}$ & -0.4595$^{***}$ & -0.6199$^{***}$      & -0.0446\\   
                        & (0.1761)        & (0.0420)        & (0.0543)             & (0.0790)\\   
   \midrule
   Observations         & 25,128,520      & 25,128,520      & 16,357,156           & 16,357,156\\  
   F-test (1st stage) & 111,242.0       & 111,242.0       & 135,873.3            & 135,873.3\\  
\midrule \midrule
\multicolumn{5}{l}{\emph{Clustered (listener level) standard-errors in parentheses}}\\
\multicolumn{5}{l}{\emph{All regressions include listener and period fixed effects}}\\
\multicolumn{5}{l}{\emph{Signif. Codes: ***: 0.01, **: 0.05, *: 0.1}}\\
\end{tabular}
\end{table}

\end{document}